\documentclass{article}

    \usepackage[preprint,nonatbib]{neurips_2024}

\usepackage[utf8]{inputenc} % allow utf-8 input
\usepackage[T1]{fontenc}    % use 8-bit T1 fonts
\usepackage{hyperref}       % hyperlinks
\usepackage{url}            % simple URL typesetting
\usepackage{booktabs}       % professional-quality tables
\usepackage{amsfonts}       % blackboard math symbols
\usepackage{nicefrac}       % compact symbols for 1/2, etc.
\usepackage{microtype}      % microtypography
\usepackage{xcolor}         % colors

\usepackage{graphicx}
\usepackage{booktabs}
\usepackage{subfigure}
\usepackage{diagbox}
\usepackage{multirow}
\usepackage{utfsym}

\title{Copiloting Diagnosis of Autism in Real Clinical Scenarios via LLMs}

\author
{Yi Jiang, \, Qingyang Shen, \, Shuzhong Lai, \, Shunyu Qi \\ \\
    \textbf{Qian Zheng\thanks{{ }Correspondence to: \texttt{\{qianzheng,lin.yao\}@zju.edu.cn}} \,, \,Lin Yao$^*$,\, Yueming Wang,\, Gang Pan} \\ \\
    Zhejiang University 
}

\begin{document}

\maketitle

\begin{abstract}
  Autism spectrum disorder(ASD) is a pervasive developmental disorder that significantly impacts the daily functioning and social participation of individuals.   Despite the abundance of research focused on supporting the clinical diagnosis of ASD,  there is still a lack of systematic and comprehensive exploration in the field of methods based on Large Language Models (LLMs), particularly regarding the real-world clinical diagnostic scenarios based on Autism Diagnostic Observation Schedule, Second Edition (ADOS-2).   Therefore, we have proposed a framework called ADOS-Copilot, which strikes a balance between scoring and explanation and explored the factors that influence the performance of LLMs in this task.   The experimental results indicate that our proposed framework is competitive with the diagnostic results of clinicians, with a minimum MAE of 0.4643, binary classification F1-score of 81.79\%, and ternary classification F1-score of 78.37\%.  Furthermore, we have systematically elucidated the strengths and limitations of current LLMs in this task from the perspectives of ADOS-2, LLMs' capabilities, language, and model scale aiming to inspire and guide the future application of LLMs in a broader fields of mental health disorders. We hope for more research to be transferred into real clinical practice, opening a window of kindness to the world for eccentric children.
  \end{abstract}
  
  \section{Introduction}
  Autism spectrum disorder (ASD) is a pervasive neurodevelopmental disorder characterized by restricted social communication, repetitive behaviors, and specific interests. It significantly impacts the quality of life and well-being of individuals affected~\cite{hirota2023autism}. According to a 2021 report from the Centers for Disease Control and Prevention (CDC) in the United States, the prevalence of ASD among children under the age of eight is approximately 1 in 44~\cite{maenner2021prevalence}. However, research indicates that the neural impairments in individuals can be early detected and timely intervention have been shown to improve language abilities and behavioral habits in children with autism, enabling them to reintegrate into society successfully. Currently, clinical diagnosis primarily relies on the Autism Diagnostic Observation Schedule, Second Edition (ADOS-2)~\cite{molloy2011use}, and heavily depends on the professional expertise of doctors. However, due to limitations in the number of qualified professionals and the inherent individual differences among children, diagnostic errors and missed cases are inevitable~\cite{hayes2022autism}, causing children to miss the optimal treatment window.
  
  With the advancement of computer technology, numerous computer-assisted methods for ASD diagnosis have emerged. One category involves simplifying clinical paradigms and collecting data from various modalities such as electroencephalography (EEG)~\cite{han2022multimodal}, eye-tracking (ET)~\cite{wei2019saliency}, video~\cite{liu2023assessing} or multi-modal~\cite{chen2019attention, li2021two} for diagnostic purposes. However, due to the simplification of real clinical paradigms, their clinical utility is significantly compromised. On the other hand, another category involves utilizing authentic paradigms and conducting automatic assessments by directly collecting data, such as speech~\cite{sadiq2019deep,song2023multimodal}, from real diagnostic. Their limitations lie in the fact that the method operates as a black box, lacking sufficient interpretability, and therefore cannot effectively assist doctors in decision-making.
  
  Recently, the remarkable capabilities of LLMs have sparked a research frenzy. Their unique ability to comprehend and reason with long contexts, coupled with the powerful knowledge base acquired through pre-training, allows them to better adapt to downstream tasks even without fine-tuning. This has led to the increasing utilization of LLMs in various diagnostic scenarios~\cite{perlis2024clinical, liu2023assessing, hu2024exploiting}, with significant improvements in their effectiveness. Therefore, we aim to input interactive dialogue texts between child and doctor from the real clinic diagnosis into LLMs to enable the generation of score and explanation for corresponding items for diagnosis. But base on our preliminary experiments, we must acknowledge that the current large-language models are not yet capable of directly handling this task. Therefore, our paper focuses on three main aspects:
  \begin{itemize}
      \item We proposed an evaluation framework called \textbf{ADOS-Copilot}, which leverages the techniques of \textbf{In-context Enhancement}, \textbf{Interpretability Augmentation} and \textbf{Adaptive Fusion} to address the limitations of LLMs in ASD diagnosis task based on real ADOS-2 clinic scenario.
      \item We conducted a comprehensive analysis of the reasons why existing mainstream large-language models are unable to perform well in ADOS-2 clinical diagnosis tasks.
      \item Extensive experimental results and analysis demonstrated that our proposed framework is competitive with doctors' diagnosis result and provides detail evidence to support the assessment, thereby assisting doctors in making more accurate and objective diagnoses and treatments.
  \end{itemize}
  
  \section{Related work}
  \subsection{Computer-aided ASD diagnosis}
  % To facilitate easier access to knowledge about the clinical diagnosis of ASD, researchers have established a Chinese knowledge base for early ASD screening called AsdKB ~\cite{wu2023asdkb}. 
  % To address the various atypical behaviors associated with ASD, researchers have extensively explored biomarkers for automatic ASD diagnosis using computer technology. Compared to EEG ~\cite{han2022multimodal,bosl2018eeg} and fMRI ~\cite{santana2022rs}, ET~\cite{frazier2018development,wei2019saliency,xia2023identification,fang2020identifying} is a more cost-effective and convenient method widely used in ASD diagnosis. 
  % In addition to using individual eye movement features, 
  To develop automated diagnosis of ASD using computer technology, researchers have extensively investigated atypical behavioral features of ASD through methods like EEG ~\cite{han2022multimodal,bosl2018eeg}, fMRI ~\cite{santana2022rs}, and ET ~\cite{frazier2018development,wei2019saliency,xia2023identification,fang2020identifying}.
  ~\cite{han2022multimodal} proposed a multimodal diagnostic framework for identifying ASD by combining EEG and ET data. ~\cite{drimalla2020towards} designed a simulated interaction task that uses a standard 7-minute simulated dialog via video to assess multiple biomarkers of social interaction deficits, including gaze behavior, facial expressions, and voice characteristics. This approach specifically targets joint attention (JA) impairments, emotional disorders, and language impairments in ASD. 
  % ~\cite{chen2019attention} employed a photo-taking task where subjects freely explore the environment in a more ecological setting, utilizing temporal information in eye movements while viewing images for the first time. 
  Acoustic and text-based features have also shown relevance in assessing children’s language and communicative behaviors for ASD diagnosis ~\cite{probol2024autism,ashwini2023spasht}. ~\cite{adilakshmi2023medical} explored the semantic and pragmatic language features in children to understand their significance in diagnosing ASD. To overcome the requirement for specialized professionals and extensive resources, some studies have adopted a natural language processing approach using electronic health records ~\cite{rubio2024enhancing} and even online text ~\cite{kim2022asdclaims,chen2023enhancing}.
  \subsection{ADOS-2 with machine learning}
  Although these studies above have made considerable progress in ASD diagnosis, translating research findings from lab settings back into clinical practice remains challenging. To address this, researchers extended their studies to the clinical environment of ADOS-2, for more details about ADOS-2 see Appendix ~\ref{sec:ados-2}, despite the difficulties in data acquisition. 
  % The ADOS-2 consists of a series of structured and semi-structured tasks, typically taking 40-60 minutes to administer.For more details about ADOS-2, see Appendix B. 
  Various paradigms based on ADOS-2 have been proposed for feature extraction and diagnosis. For instance, ~\cite{liu2023assessing} assessed ASD language, cognition, and attention by transforming ADOS-2 evaluation tasks into 9 social skill task scenarios, utilizing image-language pre-training models for score prediction. 
  % Similarly, ~\cite{liu2023social} used a multimodal framework to achieve social recognition of JA through vision-based human behavior perception. 
  % ~\cite{song2023multimodal} combined automatic name detection with pose tracking and head pose estimation in computer vision to evaluate the response to name in children with ASD. 
  Based on the ADOS-2 process, ~\cite{cheng2023computer} designed and proposed a standardized platform for stimulating, gathering, analyzing, modeling, and interpreting human behavioral data for ASD diagnosis. Additionally, ~\cite{cheng2023computer,macfarlane2022combining,ochi2019quantification,sadiq2019deep} directly used speech data from the ADOS-2 clinical assessment process for ASD prediction. Among these, ~\cite{macfarlane2022combining} used speech data from the entire process, while ~\cite{sadiq2019deep} predicted ADOS-2 Calibrated Severity Scores (CSS).
  
  \subsection{LLMs in healthcare}
  Although there are still some doubts regarding the application of large language models (LLMs) in the medical field, research has been extensively conducted in this area ~\cite{dergaa2024chatgpt,guo2024large,hua2024large,zhou2023survey,mehandru2024evaluating}. LLMs have demonstrated promising contextual understanding and zero-shot and few-shot capabilities for many medical scales and subjective descriptions. 
  Studies ~\cite{perlis2024clinical,elyoseph2024assessing,abbasian2024knowledge,lim2023artificial} used case vignettes to evaluate the differences between LLM diagnoses and expert diagnoses in areas such as depression ~\cite{perlis2024clinical,elyoseph2024assessing} and diabetes ~\cite{abbasian2024knowledge}. 
  % The results showed that LLMs, enhanced with prompt engineering techniques, have significant potential. 
  % Based on natural language dialogue with the user, ~\cite{qin2023read} applied LLMs to depression diagnosis, not only providing a diagnosis but also diagnostic evidence and personalized recommendations. 
  Studies ~\cite{xu2024mental,kim2024health} compared the performance of various LLMs on health tasks. ~\cite{singhal2023large} introduced a benchmark that combines 6 existing medical question-answering datasets spanning professional medicine, research, consumer queries, and a new dataset of medical questions searched online. 
  In the field of ASD, LLMs have effectively aided in diagnosis ~\cite{hu2024exploiting}, treatment ~\cite{cho2023evaluating}, and daily life management ~\cite{jang2024s}. In ~\cite{hu2024exploiting}, researchers used LLMs to perform binary classification on the scores of ASD adults executing the ADOS-2 A4 task, achieving an accuracy rate of 81.82\%. They also used LLMs to analyze 10 social language impairment features of ASD.
  
  Our study utilized the complete range of ADOS-2 clinical voice data and advanced the individual scoring of ADOS-2 through In-context Enhancement, Interpretability Augmentation, and Adaptive Fusion Method to propose the framework \textbf{ADOS-Copilot}. 
  % This approach aimed to predict ADOS-2 items, thereby enhancing the clinical relevance and utility of the results. 
  Our method can reduce healthcare costs, including patient testing fees and physician training expenses, and  minimize the potential for subjectivity in human diagnoses. The explainability of LLMs also contributes to the objectivity of results, aiding patients in understanding their condition and facilitating targeted treatment. However, it is important to acknowledge that the use of LLMs raises privacy and ethical concerns. Ensuring adherence to relevant ethical codes by LLM providers and further testing voice-based de-identification methods is essential. Furthermore, increased reliance on technology by doctors may lead to a higher risk of misdiagnosis.
  
  % \usepackage{booktabs}
  % \usepackage{graphicx} 
  \begin{table}[]
  \caption{Comparison of our work with other ADOS-2-based approaches. Our work leverages the full spectrum of speech data from the ADOS-2 M3 clinical evaluation, allowing for automated scoring of all language-related items, with enhanced accuracy achieved through a combination of large language models and rule-based methods.}
  \label{tab:com}
  \resizebox{\columnwidth}{!}{%
  \begin{tabular}{@{}lllllllllllll@{}}
  \toprule
  Works &
    Participant &
    Paradigm &
    Clinical Data &
    Used ADOS-2 Tasks &
    Data Type &
    Multimodal &
    LLM &
    ML &
    Explainability &
    ASD Diagnosis &
    ADOS-2 item scoring &
    CSS scoring \\ \midrule
  Liu~\cite{liu2023assessing} &
    Children &
    Based on ADOS-2 &
    \usym{2715} &
    NA &
    Video &
    \usym{1F5F8} &
    \usym{2715} &
    \usym{1F5F8} &
    \usym{2715} &
    \usym{2715} &
    NA &
    NA \\
  Cheng~\cite{cheng2023computer} &
    Children &
    Bsaed on ADOS-2 &
    \usym{2715} &
    NA &
    Video &
    \usym{1F5F8} &
    \usym{2715} &
    \usym{1F5F8} &
    \usym{1F5F8} &
    \usym{1F5F8} &
    NA &
    NA \\
  MacFarlane~\cite{macfarlane2022combining} &
    Children &
    ADOS-2 &
    \usym{1F5F8} &
    1 &
    Audio &
    \usym{2715} &
    \usym{2715} &
    \usym{1F5F8} &
    \usym{1F5F8} &
    \usym{1F5F8} &
    \usym{2715} &
    \usym{2715} \\
  Ahn~\cite{ahn2023objective} &
    Children &
    ADOS-2 &
    \usym{1F5F8} &
    All &
    Video &
    \usym{2715} &
    \usym{2715} &
    \usym{1F5F8} &
    \usym{2715} &
    \usym{1F5F8} &
    \usym{2715} &
    Regression \\
  Sadiq~\cite{sadiq2019deep} &
    Children &
    ADOS-2 &
    \usym{1F5F8} &
    All &
    Audio &
    \usym{2715} &
    \usym{2715} &
    \usym{1F5F8} &
    \usym{2715} &
    \usym{1F5F8} &
    \usym{2715} &
    Regression \\
  Hu~\cite{hu2024exploiting} &
    Adults &
    ADOS-2 &
    \usym{1F5F8} &
    11/15 &
    Audio &
    \usym{2715} &
    \usym{1F5F8} &
    \usym{2715} &
    \usym{2715} &
    \usym{2715} &
    Binary Classification &
    \usym{2715} \\
  Our Work &
    Children &
    ADOS-2 &
    \usym{1F5F8} &
    All &
    Audio &
    \usym{2715} &
    \usym{1F5F8} &
    \usym{1F5F8} &
    \usym{1F5F8} &
    \usym{1F5F8} &
    Ternary Classification &
    Ternary Classification \\ \bottomrule
  \end{tabular}%
  }
  \end{table}
  
  \begin{figure}
      \centering
      \includegraphics[width=0.9\linewidth]{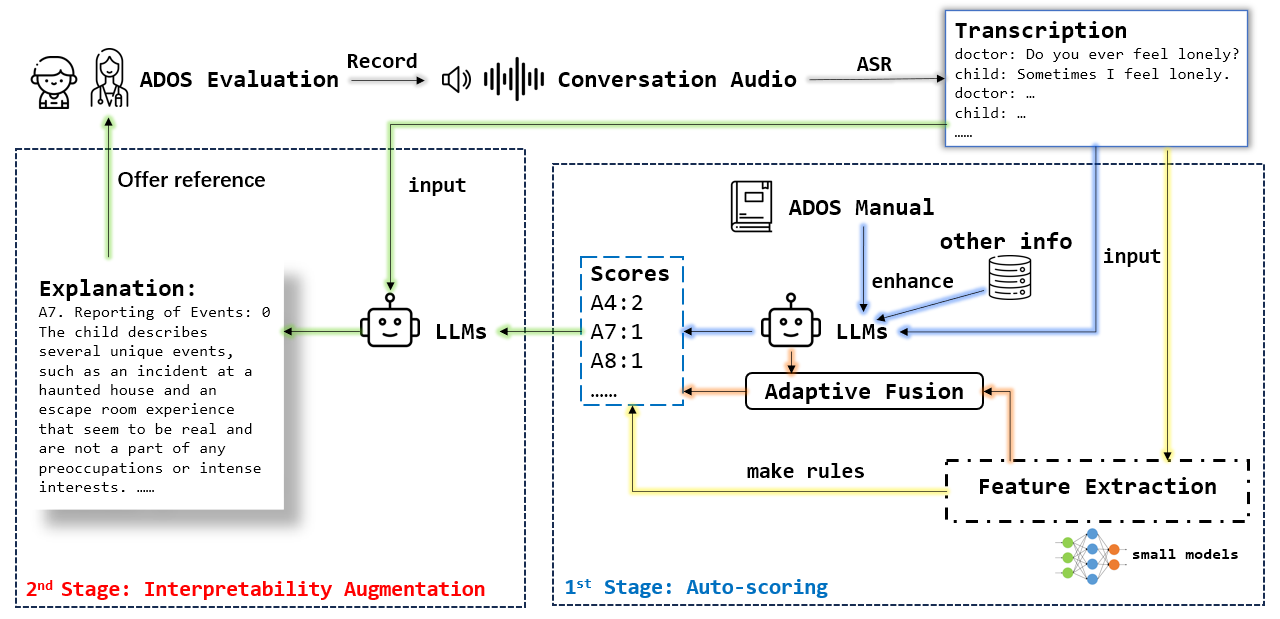}
     \caption{Overall pipeline of our framework. Firstly, we preprocess and transcribe the complete long dialogue text based on the original recordings from clinical ASD diagnoses. Then, we design prompts for In-context Enhancement, using the long text dialogue as input for LLMs to generate scores and justifications for the eight items used in the ADOS clinical diagnoses: A4, A7, A8, B4, B7, B9, B10, and B11. Furthermore, we also feed the long dialogue text into a rule-based method to extract features and then integrated with the results from LLMs, combining the strengths of both approaches to achieve the best scoring performance. Finally, in order to delve deeper into the meaning behind the scores and assist in clinical decision-making, we further input the output fusion results and the original long dialogue text into LLMs to obtain more detailed textual support and additional explanatory evaluations via Interpretability Augmentation in second stage.}
      \label{fig:pipeline}
  \end{figure}
  
  \section{Methodology}
  Our proposed framework is shown in Figure ~\ref{fig:pipeline}. In concisely, we obtained scores for the 8 items of the ADOS-2 Module 3(M3) language section using the \textbf{In-context Enhancement} and \textbf{Adaptive Fusion} method. Then, we utilized \textbf{Interpretability Augmentation} to generate more detailed explanations. It is worth mentioning that ADOS-2 M3 has a total of 14 clinical diagnostic items, with the excluded items being purely non-verbal assessments.
  
  \subsection{In-context enhancement prompt}
  We introduced the technique of \textbf{in-context enhancement} to address the challenge of current LLMs being unable to utilize generic prompt templates for solving task of ADOS-2 scoring and explanation in real clinical scenarios.
  
  In-context Enhancement Prompt details are shown in Appendix ~\ref{sec:in-context} with the Figure ~\ref{fig:base prompt}. We aim to design such a prompt that enables the model to focus on scoring while providing corresponding justifications simultaneously called \textbf{Scoring\&Explanation}(zero-shot). We design such prompt based on the following assumptions: reasoning is aimed at better scoring, and scoring is aimed at better reasoning. This assumption aligns with intuition: if the task focuses solely on scoring without providing justifications, the model may resort to tricks, become lazy in thinking, and consequently yield poorer scoring results. Conversely, if the emphasis is solely on reasoning without scoring, there won't be a reasonable standard for the reasoning process, resulting in lower interpretability of the justifications provided~\cite{yang2023towards}. It is worth mentioning that in Figure ~\ref{fig:base prompt}, the \textcolor{blue}{\(-(optional)\)} prefix text can be replaced with more settings called Only-Scoring and Scoring\&Explanation(few-shot). More zero or few shot setting detail and experiment will be discussed in Appendix ~\ref{sec:zs vs fs}. 
  
  By leveraging the contextual learning ability~\cite{brown2020language}, LLMs are capable of completing the various task without updating parameters. Therefore, we introduce the criteria and procedures of the ADOS-2 as well as the prior statistical infomation as in-context enhancement prompts. For criteria, some researches have shown that LLMs tend to make more errors when confronted with more complex conversational contexts~\cite{bhattacharya2023context,li2023compressing,xu2024mental}. Therefore, we introduce more detailed criteria to guide LLMs in properly activating existing clinical knowledge to better understand this complex task. Inspired by ~\cite{qin2023read}, we incorporate the scoring criteria of the eight items used in clinical ASD diagnosis into the system prompt.
  For procedures, we consider that the evaluation of ADOS-2 consists of a series of consecutive scenario-based interactions. However, it is challenging to segment the individual scenes during the process of speech recognition. In order to facilitate the model's understanding of the evaluation scenarios in ADOS-2, we include excerpts from the assessment process manual used as a reference in hospitals as part of the prompt. For prior information, to better align with the assessment results of doctors and mitigate potential biases inherent in LLMs, we introduce prior statistical information into the prompt. The statistics are collected by hospital, including the mean of each item and the proportion of ASD and TD children. 
  
  \subsection{Adaptive integrating LLMs with rule-based model for enhanced ASD diagnosis}
  This section introduces an advanced, hybrid methodology that merges the data-driven rule-based model, aligning with ADOS-2 criteria, with the adaptability of LLMs. Our aim is to establish a more comprehensive and reliable diagnostic system for Autism Spectrum Disorder (ASD), addressing the lack of objective and quantitative tools in current ASD diagnosis.
  
  We presents a data-driven rule-based model, addressing the need for objective, quantitative diagnosis tools. It extracts tailored features from child's communication and social interactions linked to ADOS-2 and clinical guidelines for quantifying~\cite{ashwini2023spasht, ochi2019quantification}. Subsequently, rule-based scoring is applied, translating these features into quantitative assessments for individual ADOS-2 sub-items. Readers are encouraged to refer to Appendix ~\ref{sec:paradigm design}, which provides supplementary information on the design frameworks underlying our rule-based scoring mechanism.
  
  To augment the robustness and validity of scoring framework, we integrate LLMs with rule-based model through an adaptive fusion strategy. This pioneering fusion recognizes the complementary strengths of LLMs and rule-based systems, optimizing performance of our scoring system. Central to this integration is a adaptive weighting schema, which assigns variable importance to each model’s predictions based on Mean Absolute Error (MAE) assessments. Consequently, the model that demonstrates superior performance in real-world scenarios is given greater influence, enhancing overall accuracy and adaptability to diverse data intricacies and contexts.

  In conclusion, we presents an approach directly addressing the challenge of enhancing prediction reliability across diverse scenarios where traditional single model fall short, through a data-driven balancing act between LLMs and rule-based systems. The fused prediction score \(\hat{y}_i\) is expressed as:
  \begin{equation}
  \hat{y}_i = {\alpha_{LLM}}_i \cdot {f_{LLM}}_i(x) + {\alpha_{Rule}}_i \cdot {f_{Rule}}_i(x) \; .
  \end{equation}
  Here, \(\alpha_{LLM}\) and \(\alpha_{Rule}\) denote the adaptive weights allocated to the LLM and rule-based predictions respectively for the different ADOS-2 items, \({f_{LLM}}_i(x)\) and \({f_{Rule}}_i(x)\) denote the predictions of the LLM and rule-based. Specifically, the calculation of these coefficients is expressed as:
  \begin{equation}
  {\alpha_{LLM}}_i = \frac{\exp(-{MAE_{LLM}}_i)}{\exp(-{MAE_{LLM}}_i) + \exp(-{MAE_{Rule}}_i)}, \quad {\alpha_{Rule}}_i = 1 - {\alpha_{LLM}}_i \; .
  \end{equation}
  
  % \begin{figure}
  %     \centering
  %     \includegraphics[width=0.9\linewidth]{figures/image.png}
  %     \caption{In-context Enhancement prompt for ASD-Eval. Where \(Prompt_{criteria}\) refers to the prompt of clinical ADOS criteria, \(Prompt_{m3}\) refers to the prompt of clinical ADOS-2 Module 3 procedures, \(Prompt_{stat}\) refers to the prior information of the ASD and TD children, \(Transcript_i\) refers to the pro-processing dialogue texts between doctor and child.}
  %     \label{fig:enter-label}
  % \end{figure}
  
  \subsection{Interpretability augmentation via the second stage}
  Based on our experimental results, we find that under the prompt setting of Scoring\&Explanation, all LLMs perform better in terms of scoring results, including MAE and classification evaluation metrics, in the zero-shot setting compared to the few-shot or Only-Scoring setting. This suggests that the few-shot setting may primarily provide a format guideline and allow LLMs to extract more task-related information from the format guidelines~\cite{xie2021explanation}, thereby appearing a more professional illusion. However, we aim to provide more reasonable justifications and evidence while ensuring accurate scoring. Therefore, we need a method that can achieve a balance between the two objectives. So partially inspired by ~\cite{luo2024unlocking}, we introduce a second stage to
  enhance the interpretability: in the first stage, we utilize the zero-shot setting to perform the initial scoring and evidence search, aiming to obtain the best scores. In the second stage, we use the scores obtained from the first stage and the original dialogue text as the user prompt. Based on this prompt, we employ a reasoning chain to guide the LLMs to directly extract the most relevant excerpts from the original dialogue text that correspond to the scores and generate the output accordingly. This two-stage approach allows us to achieve both accurate scoring and provide more relevant justifications based on the original dialogue text. The prompt's detail using for second stage is shown in Appendix ~\ref{sec:second stage prompt} with the Figure ~\ref{fig:two-stage-prompt}.

  % \subsubsection{Paradigm 1}
  % \subsubsection{Paradigm 2}
  % \subsubsection{Paradigm 3}
  \section{Experiments and results}
  \subsection{Dataset}
  The dataset utilized in this study was recorded in a clinical context, consisting of 28 audio samples of the whole ADOS-2 assessment processes. We then transcribe the audio into text. Written informed consent was obtained from the parents or caregivers of all participants prior to the assessment. See more details in Appendix ~\ref{sec:dataset}.
  \subsection{Experimental setup}
  \paragraph{LLMs}
  We explore a wide span of LLMs in our experiment, including GPT-4-turbo~\cite{achiam2023gpt}, Gemini1.5-Pro~\cite{reid2024gemini}, Claude3, Llama3-8b~\cite{llama3modelcard}, Mixtral~\cite{jiang2023mistral, jiang2024mixtral}, Qwen1.5~\cite{bai2023qwen}, Glm~\cite{du2022glm}, Yi-34b~\cite{young2024yi}, Kimi. It is worth noting that our data is derived from a Chinese dataset. Therefore, we have opted to use more LLMs that have been trained on a larger proportion of Chinese pretraining data. Additionally, some models were not trained using Chinese word embeddings, consequently, during experimentation, we will translate both the data and prompts into English when using them, which may lead to inevitable accuracy loss.  All LLMs use the default temperature set by their creators.
  
  \paragraph{Metrics}
  We use MAE to measure the performance of each 8 items (A4,A7,A8,B4,B7,B9,B10,B11) like many previous works ~\cite{cheng2023computer}, since the magnitude of these scores reflect the degree that the item aims to evaluate. We also calculate the mean MAE of the 8 items to reflect the overall performances. Besides, accounting that the ultimate goal of ADOS-2 is diagnosis, some classification metrics are included. The classification results are calculated using the items scored by model combined with 6 remianing items directly copied from doctor's scoring. The classification task can be either binary (Autism Spectrum Disorder and Non-Spectrum Disorder) or ternary (Autism, Autism Spectrum Disorder, and Non-Spectrum Disorder). The scoring criteria align with clinical judgment, where the overall judgment is based on the sum of scores from the 14 items for classification.
  
  \subsection{Results}
  \begin{table}[]
  \caption{In-context enhancement ablation result in Qwen1.5-32b, where \textit{Concise} refers to a simplified version of the criteria, retaining only the names of the scoring items and the scoring ranges while excluding specific scoring details (Appendix ~\ref{sec:concise} for detailed comparison), \textit{C} refers to the standard criteria, \textit{M} refer to the ADOS-2-M3 clinical diagnostic procedures and \textit{S} refers to the prior statistical information.}
  \label{tab:ablation}
  \resizebox{\textwidth}{!}{%
  \begin{tabular}{@{}llllllllllllllll@{}}
  \toprule
  \multicolumn{1}{c}{\textbf{prompt}} &
    \multicolumn{1}{c}{\textbf{A4}} &
    \multicolumn{1}{c}{\textbf{A7}} &
    \multicolumn{1}{c}{\textbf{A8}} &
    \multicolumn{1}{c}{\textbf{B4}} &
    \multicolumn{1}{c}{\textbf{B7}} &
    \multicolumn{1}{c}{\textbf{B9}} &
    \multicolumn{1}{c}{\textbf{B10}} &
    \multicolumn{1}{c}{\textbf{B11}} &
    \multicolumn{1}{c}{\textbf{avg}} &
    \multicolumn{1}{c}{\textbf{2-acc}} &
    \multicolumn{1}{c}{\textbf{2-precision}} &
    \multicolumn{1}{c}{\textbf{2-f1}} &
    \multicolumn{1}{c}{\textbf{3-acc}} &
    \multicolumn{1}{c}{\textbf{3-precision}} &
    \multicolumn{1}{c}{\textbf{3-f1}} \\ \midrule
    Random & 0.8934 & 0.7739 & 0.8566 & 0.9879 & 0.8217 & 0.8095 & 0.8568 & 0.8813 & 0.8601 & 0.5927 & 0.6451 & 0.5167 & 0.4493 & 0.5534 & 0.4037 \\
  Concise &
    0.9643 &
    0.4286 &
    0.5714 &
    0.8214 &
    0.6071 &
    0.4643 &
    0.6429 &
    0.7500 &
    0.6266 &
    0.5714 &
    0.5714 &
    0.5429 &
    0.4286 &
    0.5533 &
    0.4667 \\
  C &
    \textbf{0.6071} &
    0.4643 &
    0.5714 &
    0.5357 &
    0.4643 &
    0.4286 &
    0.5714 &
    0.6429 &
    0.5227 &
    0.6071 &
    0.7734 &
    0.5158 &
    0.5000 &
    0.7384 &
    0.4556 \\
  C+M &
    0.7143 &
    0.6429 &
    0.6071 &
    0.7500 &
    0.5000 &
    0.4286 &
    0.6071 &
    0.6429 &
    0.5747 &
    0.6786 &
    0.7991 &
    0.6306 &
    0.5000 &
    0.7173 &
    0.5107 \\
  C+S &
    0.7857 &
    \textbf{0.3214} &
    0.5714 &
    0.2857 &
    \textbf{0.4286} &
    0.6071 &
    0.5000 &
    0.6786 &
    0.4870 &
    0.7500 &
    0.7500 &
    \textbf{0.7490} &
    0.6071 &
    0.6625 &
    0.6336 \\
  C+M+S &
    1.0000 &
    0.4643 &
    \textbf{0.4643} &
    \textbf{0.2857} &
    0.5000 &
    0.4643 &
    \textbf{0.4643} &
    \textbf{0.6429} &
    \textbf{0.4805} &
    \textbf{0.7500} &
    \textbf{0.7813} &
    0.7381 &
    \textbf{0.6786} &
    \textbf{0.7753} &
    \textbf{0.6969} \\
  % P+concise &
  %   0.9643 &
  %   0.6786 &
  %   0.6071 &
  %   0.7500 &
  %   0.4643 &
  %   0.4643 &
  %   0.5357 &
  %   0.7143 &
  %   0.5844 &
  %   0.7143 &
  %   0.7158 &
  %   0.7113 &
  %   0.6071 &
  %   0.6789 &
  %   0.6321 \\
  % P &
  %   0.6429 &
  %   0.3929 &
  %   0.5714 &
  %   0.6429 &
  %   0.5000 &
  %   \textbf{0.3929} &
  %   0.5714 &
  %   \textbf{0.6429} &
  %   0.5422 &
  %   0.5357 &
  %   0.2870 &
  %   0.3738 &
  %   0.4286 &
  %   0.2839 &
  %   0.3358 \\
  % P+D &
  %   0.6429 &
  %   0.4643 &
  %   0.5714 &
  %   0.6071 &
  %   0.4643 &
  %   0.4286 &
  %   0.5714 &
  %   0.6786 &
  %   0.5357 &
  %   0.6071 &
  %   0.7733 &
  %   0.5158 &
  %   0.4643 &
  %   0.7406 &
  %   0.4424 \\
  % P+S &
  %   1.0000 &
  %   0.5000 &
  %   0.5714 &
  %   0.1786 &
  %   0.4643 &
  %   0.5714 &
  %   0.5714 &
  %   \textbf{0.6429} &
  %   0.5357 &
  %   0.7143 &
  %   0.7277 &
  %   0.7051 &
  %   0.5357 &
  %   0.7215 &
  %   0.5770 \\
  % P+D+S &
  %   0.8214 &
  %   0.4643 &
  %   0.5000 &
  %   \textbf{0.1250} &
  %   \textbf{0.3214} &
  %   0.6071 &
  %   0.6071 &
  %   0.6786 &
  %   \textbf{0.4757} &
  %   \textbf{0.8214} &
  %   \textbf{0.8345} &
  %   \textbf{0.8179} &
  %   \textbf{0.7143} &
  %   \textbf{0.8627} &
  %   \textbf{0.7451} \\ 
    \bottomrule
  \end{tabular}%
  }
  \end{table}

  \paragraph{Validation for in-context enhancement prompt strategy}
  Based on the excellent performance of the Qwen1.5 series, we conducted detailed prompt ablation experiments on the 32b model shown in Table ~\ref{tab:ablation}. The results showed that the prompt consisting of standard ADOS-2 scoring criteria, the diagnostic procedures of ADOS-2, and prior information(\textit{C+M+S}) which is described in Figure ~\ref{fig:base prompt} yielded the best results. This prompt design effectively maximized the potential of LLMs and enabled them to successfully tackle this complex task. Compared to the concise setting, the results of the in-context enhancement prompt show a decrease of 14.61\% in average MAE across the eight items. Additionally, there is an increase of 17.86\% in binary classification accuracy, 19.52\% in F1 score, 25.00\% in ternary classification accuracy, and 23.02\% in F1 score which clearly demonstrates that utilizing in-context enhancement prompts can effectively tap into the potential of LLMs in this task.
  
  Additionally, we observed that different prompt settings can indirectly and significantly impact the accuracy of certain item, thereby influencing the MAE and classification. For standard criteria, it is crucial for LLMs to include it to understand more objectives and justifications of each scoring criterion, allowing for more objective and professional scoring. We can observe significant improvements in all performance metrics when switching from concise to standard criteria. For the knowledge of ADOS-2 procedure, it could help models to understand the scenario of the conversation with incorporating domain-specific knowledge into the models. For the prior statistical information, it is clearly helpful to lowering B4's MAE. The statistics of ADOS-2 scoring could guide LLMs to give more suitable scores. For example, if mean of an item is pretty low, the model will be more hesitant to give a score of 2. In this sense, the prior information could help to align models'pattern to humans'.
  
  \paragraph{Validation for rule-based model and fusion strategy}
  By scrutinizing the entries within the 'rule' row of Table ~\ref{tab:fusion-result}, it becomes evident that the rule-based model excels in the majority of the evaluated criteria, thereby attesting to its proficient performance. Of particular note, items A4 and B4 yield MAE scores that are less than LLM's. This notable accomplishment can be attributed to the meticulous quantification of ADOS-2 alongside the adept utilization of sentiment analysis techniques, underscoring the efficacy of these methodologies in enhancing predictive accuracy and precision. 
  
  Table ~\ref{tab:fusion-result} showcases the efficacy of combining LLMs with rule-based systems. The Fusion model, by merging the two, achieves superior outcomes: compared to the LLMs, fusion model show a decrease of 2.38\% in average MAE across the eight items. In addition, there is an increase of 5.00\% in ternary classification accuracy, This evidently illustrates that employing a fusion strategy can efficiently harness the capabilities of both LLMs and rule-based models for this particular task. Thus, Fusion models improve overall performance and robustness, affirming the merit in hybrid solutions. For more detailed results of rule-based model and fusion strategy, please refer to Appendix ~\ref{sec:more-results}.
  
  \begin{table}[]
  \caption{Comparative result of LLM, rule-based system, and their fusion. Among these, LLM refers to the LLM with the best performance, qwen1.5-72b; Rule denotes the rule-based model; and Fusion represents the model that integrates both.}
  \label{tab:fusion-result}
  \resizebox{\textwidth}{!}{%
  \begin{tabular}{@{}llllllllllllllll@{}}
  \toprule
  \multicolumn{1}{c}{\textbf{model}} &
    \multicolumn{1}{c}{\textbf{A4}} &
    \multicolumn{1}{c}{\textbf{A7}} &
    \multicolumn{1}{c}{\textbf{A8}} &
    \multicolumn{1}{c}{\textbf{B4}} &
    \multicolumn{1}{c}{\textbf{B7}} &
    \multicolumn{1}{c}{\textbf{B9}} &
    \multicolumn{1}{c}{\textbf{B10}} &
    \multicolumn{1}{c}{\textbf{B11}} &
    \multicolumn{1}{c}{\textbf{avg}} &
    \multicolumn{1}{c}{\textbf{2-acc}} &
    \multicolumn{1}{c}{\textbf{2-precision}} &
    \multicolumn{1}{c}{\textbf{2-f1}} &
    \multicolumn{1}{c}{\textbf{3-acc}} &
    \multicolumn{1}{c}{\textbf{3-precision}} &
    \multicolumn{1}{c}{\textbf{3-f1}} \\ \midrule
  LLM    & 0.8214 & \textbf{0.4643} & 0.5000 & 0.1250 & 0.3214 & 0.6071 & \textbf{0.6071} & \textbf{0.6786} & 0.4756 & \textbf{0.8214} & 0.8345 & \textbf{0.8179} & 0.7143 & \textbf{0.8627} & 0.7451 \\
  Rule   & 0.6938 & 0.5137 & 0.5190 & \textbf{0.1178} & 0.3245 & 0.6809 & 0.7196 & 0.8252 & 0.4982 & 0.7903 & 0.8271 & 0.7887 & 0.6854 & 0.7426 & 0.6833 \\
  Fusion & \textbf{0.6885} & 0.5165 & \textbf{0.4823} & 0.1877 & \textbf{0.3036} & \textbf{0.5383} & 0.6240 & 0.6951 & \textbf{0.4643} & \textbf{0.8214} & \textbf{0.8661} & 0.8129 & \textbf{0.7500} & 0.8626 & \textbf{0.7837} \\ 
  Random & 0.8934 & 0.7739 & 0.8566 & 0.9879 & 0.8217 & 0.8095 & 0.8568 & 0.8813 & 0.8601 & 0.5927 & 0.6451 & 0.5167 & 0.4493 & 0.5534 & 0.4037 \\
  \bottomrule
  \end{tabular}%
  }
  \end{table}
  
  % \usepackage{booktabs}
  % \usepackage{graphicx}

  % \begin{figure}
  %     \centering
  %     \includegraphics[width=0.75\linewidth]{figures/lang1.png}
  %     \caption{Effect of language}
  %     \label{fig:elang}
  % \end{figure}
  
  % \begin{figure}[htbp]
  % \centering
  %   \subfigure[Effect of language]{
  %     \begin{minipage}[t]{0.46\textwidth}
  %       \centering
  %       \includegraphics[width=\textwidth]{figures/lang2.png}
  %       % \label{fig:img1}
  %     \end{minipage}
  %   }
  %   \hfill
  %   \subfigure[Result for Average MAE of all LLMs for eight ADOS items. It is worth mentioning that LLMs exhibit significant variations in MAE on B4, which requires additional discussion.]{
  %     \begin{minipage}[t]{0.44\textwidth}
  %       \centering
  %       \includegraphics[width=\textwidth]{figures/mae_for_eight_item.png}
  %       % \label{fig:img2}
  %     \end{minipage}
  %   }
  %   % \caption{Effect of scale}
  %   % \label{fig:side_by_side}
  % \end{figure}

  % \begin{figure}
  %     \centering
  %     \includegraphics[width=0.5\linewidth]{figures/qwen-size}
  %     % \caption{}
  %     \label{fig:szie}
  %     \quad
  %     \includegraphics[width=0.5\linewidth]{figures/mistral-size.png}
      
  % \end{figure}
  % \begin{figure}
  %     \centering
  %     \includegraphics[width=0.5\linewidth]{figures/mistral-size.png}
  % \end{figure}
  
  % \usepackage{booktabs}
  % \usepackage{graphicx}

  \begin{figure}
      \centering
      \includegraphics[width=0.9\linewidth]{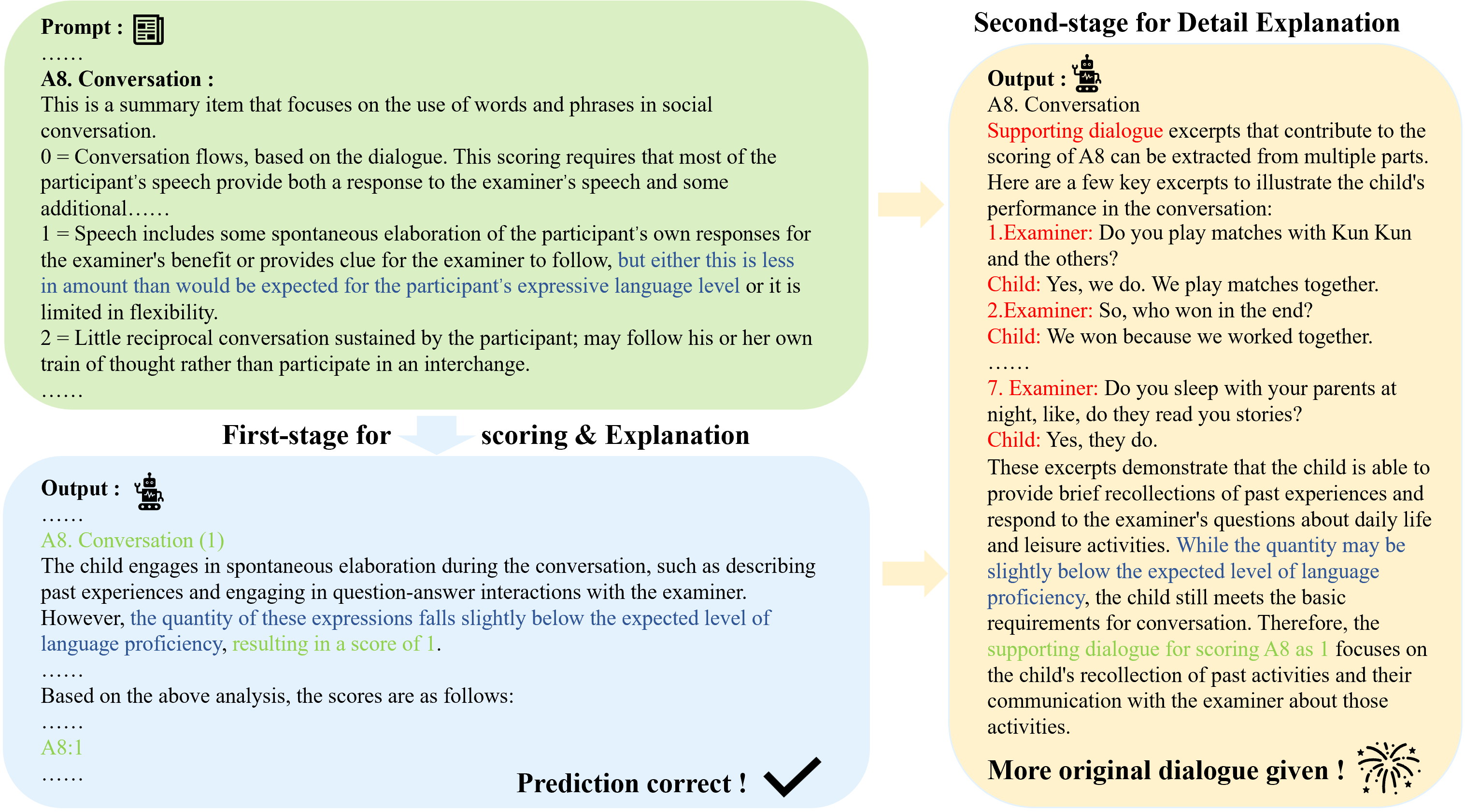}
      \caption{Case study for our framework generated by Qwen1.5-32b. The left part refers to the Scoring\&Explanation Stage, where the explanations are relatively general. The right part refers to the Interpretability Augmentation Stage, which include truncated segments from the original dialogue that support the scoring decisions made by our framework. Upon analyzing the output text, the \textcolor{blue}{blue} part refers to the references to the ADOS-2-M3 scoring criteria, the \textcolor{green}{green} part signifies scoring consistency, and the \textcolor{red}{red} part indicates several original dialogue segments that support the scoring decisions made by our framework.}
      \label{fig:two-stage}
  \end{figure}
  \paragraph{Validation for interpretability augmentation via case study}
  From the result of our experiment, we can observe that in the first stage, due to the constraint of tokens and the dual task of scoring and explanation, the explanations provided are often general summaries rather than specific dialogue excerpts. For instance, in Figure ~\ref {fig:two-stage}, when scoring the A8 dialogue item, LLMs provide a general summary stating, "Describes past experiences and engages with the doctor's questions, with a lower-than-expected level of expressive language." This leads to a score of 1. However, using Interpretability Augmentation in the second stage, several brief dialogue excerpts are extracted from the original dialogue to support this assertion, accompanied by more detailed explanations. For example, in Figure ~\ref{fig:two-stage}, seven dialogue excerpts related to daily life and leisure activities are provided from the original dialogue to support the before assertion. Furthermore, after providing the original excerpts, ADOS-Copilot responded to the detailed scoring criteria from the first stage and reaffirmed the scores from the initial phase. Therefore, our framework can achieve consistency and accuracy in reasoning while providing accurate scoring and detailed explanations. More case studies on our framework can be found in the Appendix ~\ref{sec:case study}. 
  
  \section{Discussion}
  % Based on above basic experiment result, we conducted further exploratory discussion to explore the potential of our framework.
  
  \begin{table}[]
  \caption{LLMs baseline Result. All LLMs using the prompt mentioned by Figure ~\ref{fig:base prompt}. The table below presents the single MAE and average MAE (avg) for eight items, as well as the comprehensive metrics for binary and ternary classifications based on ADOS-2 M3 clinical diagnostic criteria.}
  \label{tab:base-result}
  \resizebox{\textwidth}{!}{%
  \begin{tabular}{@{}llllllllllllllll@{}}
  \toprule
  \multicolumn{1}{c}{\textbf{model}} &
    \multicolumn{1}{c}{\textbf{A4}} &
    \multicolumn{1}{c}{\textbf{A7}} &
    \multicolumn{1}{c}{\textbf{A8}} &
    \multicolumn{1}{c}{\textbf{B4}} &
    \multicolumn{1}{c}{\textbf{B7}} &
    \multicolumn{1}{c}{\textbf{B9}} &
    \multicolumn{1}{c}{\textbf{B10}} &
    \multicolumn{1}{c}{\textbf{B11}} &
    \multicolumn{1}{c}{\textbf{avg}} &
    \multicolumn{1}{c}{\textbf{2-acc}} &
    \multicolumn{1}{c}{\textbf{2-precision}} &
    \multicolumn{1}{c}{\textbf{2-f1}} &
    \multicolumn{1}{c}{\textbf{3-acc}} &
    \multicolumn{1}{c}{\textbf{3-precision}} &
    \multicolumn{1}{c}{\textbf{3-f1}} \\ \midrule
  Yi-34b &
    0.6071 &
    0.7143 &
    0.4643 &
    0.4286 &
    0.5893 &
    0.6964 &
    0.5179 &
    0.7500 &
    0.5860 &
    0.5714 &
    0.5681 &
    0.5577 &
    0.4286 &
    0.4748 &
    0.4420 \\
  glm4 &
    0.9643 &
    0.6429 &
    \textbf{0.3929} &
    \textbf{0.1071} &
    0.6071 &
    0.6429 &
    0.6429 &
    0.7857 &
    0.5649 &
    0.6071 &
    0.6571 &
    0.5890 &
    0.5357 &
    0.6482 &
    0.5190 \\
  glm-3-turbo &
    0.8214 &
    0.5714 &
    0.5714 &
    0.4643 &
    0.6786 &
    0.7143 &
    0.7500 &
    0.7857 &
    0.6071 &
    0.6429 &
    0.6494 &
    0.6429 &
    0.5714 &
    0.6468 &
    0.5815 \\
  kimi &
    \textbf{0.5357} &
    0.3571 &
    0.6071 &
    0.9649 &
    0.4286 &
    0.3929 &
    0.6071 &
    0.6071 &
    0.5487 &
    0.5357 &
    0.2870 &
    0.3738 &
    0.4286 &
    0.2571 &
    0.3187 \\
  qwen-max &
    0.7500 &
    \textbf{0.2142} &
    0.8571 &
    0.8214 &
    0.6429 &
    0.7500 &
    0.9286 &
    \textbf{0.5000} &
    0.6396 &
    0.6071 &
    0.6607 &
    0.5485 &
    0.4643 &
    0.5665 &
    0.4348 \\
  qwen1.5-110b &
    0.7143 &
    0.4286 &
    0.5715 &
    0.6429 &
    0.4643 &
    \textbf{0.3214} &
    0.5179 &
    0.6071 &
    0.5179 &
    0.6071 &
    0.7734 &
    0.5158 &
    0.5357 &
    0.7709 &
    0.5262 \\
  qwen1.5-72b &
    0.8571 &
    0.4643 &
    0.7143 &
    0.2857 &
    \textbf{0.3929} &
    0.5714 &
    0.5000 &
    0.6071 &
    0.4968 &
    \textbf{0.7857} &
    \textbf{0.8470} &
    \textbf{0.7714} &
    \textbf{0.7143} &
    \textbf{0.8224} &
    \textbf{0.7204} \\
  qwen1.5-32b &
    1.0000 &
    0.4643 &
    0.4643 &
    0.2857 &
    0.5000 &
    0.4643 &
    \textbf{0.4643} &
    0.6429 &
    \textbf{0.4805} &
    0.7500 &
    0.7813 &
    0.7381 &
    0.6786 &
    0.7753 &
    0.6969 \\
  qwen1.5-14b &
    0.9286 &
    0.7321 &
    0.6786 &
    0.1786 &
    0.4643 &
    0.5400 &
    0.6354 &
    0.6429 &
    0.5533 &
    0.7500 &
    0.7526 &
    0.7503 &
    0.5714 &
    0.7530 &
    0.5943 \\
  \midrule
  gpt4 &
    0.8214 &
    0.7500 &
    0.8929 &
    1.1429 &
    0.7143 &
    0.4643 &
    0.7143 &
    0.6429 &
    0.7013 &
    0.6071 &
    0.7733 &
    0.5158 &
    0.4643 &
    0.6827 &
    0.3957 \\
  gemini-1.5-pro &
    0.6429 &
    0.2857 &
    0.5714 &
    1.2857 &
    0.4643 &
    0.3571 &
    0.5714 &
    0.5357 &
    0.5649 &
    0.5714 &
    0.7619 &
    0.4490 &
    0.4286 &
    0.6607 &
    0.3282 \\
  % claude-3-sonnet &
  %   0.7500 &
  %   0.7500 &
  %   0.8929 &
  %   0.9643 &
  %   1.0000 &
  %   0.7500 &
  %   0.8929 &
  %   0.8929 &
  %   0.7987 &
  %   0.5357 &
  %   0.2870 &
  %   0.3738 &
  %   0.3929 &
  %   0.1800 &
  %   0.2470 \\
  % claude-3-haiku &
  %   0.6786 &
  %   0.4643 &
  %   0.6786 &
  %   0.5357 &
  %   0.5000 &
  %   0.4286 &
  %   0.6786 &
  %   0.6786 &
  %   0.5779 &
  %   0.5714 &
  %   0.7619 &
  %   0.4490 &
  %   0.4286 &
  %   0.6803 &
  %   0.3451 \\
  claude-3-opus &
    0.5714 &
    0.3929 &
    0.8929 &
    0.6071 &
    0.6786&
    0.5714 &
    0.8214 &
    0.8214 &
    0.6558 &
    0.5000 &
    0.4333 &
    0.4063 &
    0.3929 &
    0.3615 &
    0.3199 \\
  mixtral-8x22b &
    0.6071 &
    0.6429 &
    0.7143 &
    0.5714 &
    0.7143 &
    0.8571 &
    0.9643 &
    1.0000 &
    0.6786 &
    0.6786 &
    0.6778 &
    0.6773 &
    0.5715 &
    0.5744 &
    0.5720 \\
  mixtral-8x7b &
    0.6429 &
    0.5357 &
    0.6786 &
    0.4286 &
    0.6071 &
    0.4286 &
    0.6786 &
    0.7143 &
    0.5649 &
    0.6429 &
    0.6633 &
    0.6190 &
    0.5000 &
    0.6094 &
    0.5274 \\
  mistral-7b &
    0.8214 &
    0.7857 &
    0.6071 &
    0.7500 &
    0.5714 &
    0.6071 &
    0.5714 &
    0.7143 &
    0.6526 &
    0.5357 &
    0.5298 &
    0.5265 &
    0.4286 &
    0.4821 &
    0.4519 \\
  llama-3-8b &
    0.7500 &
    0.6429 &
    0.6786 &
    0.6429 &
    0.6429 &
    0.7857 &
    0.7500 &
    0.6429 &
    0.6494 &
    0.5714 &
    0.7619 &
    0.4490 &
    0.4643 &
    0.6893 &
    0.3600 \\
  \midrule
  random &
    0.8934 &
    0.7739 &
    0.8566 &
    0.9879 &
    0.8217 &
    0.8095 &
    0.8568 &
    0.8813 &
    0.8601 &
    0.5927 &
    0.6451 &
    0.5167 &
    0.4493 &
    0.5534 &
    0.4037 \\ \bottomrule
  \end{tabular}%
  }
  \end{table}
  
  \subsection{Further experiments conducted on the LLMs baseline}
  To validate the effectiveness and generalizability of proposed framework, We conducted extensive comparative experiments on a wide range of LLMs. Due to limitations in handling long contexts, not all famous LLMs such as llama-2~\cite{touvron2023llama} were selected. From Table ~\ref{tab:base-result}, it can be observed that the Qwen1.5 series of models perform SOTA. The 72b model achieved an average MAE of 0.4805, while the 72b model excelled in classification metrics. It achieved an accuracy of 78.57\%, precision of 84.70\%, and an F1 score of 77.14\% for binary classification. For the ternary classification, it achieved an accuracy of 71.43\%, precision of 82.24\%, and an F1 score of 72.04\%. Indeed, it is surprising that GPT-4 did not achieve better results in our task compared to the renowned Qwen1.5 series. Additionally, some models even performed worse than random in certain aspects, which indicates that the complexity of our task surpasses the models'ability to comprehend the context. Despite using methods like in-context enhancement, the inherent limitations in their foundational capabilities cannot be compensated. So what effects the evaluation made by LLMs? We will delve into this question in the subsequent section.
  
  \subsection{What effects the evaluation made by LLMs?}
  
  \begin{table}[]
  \caption{Average MAE of all LLMs for single scoring criteria. The \textcolor{blue}{blue} font refers to the item LLMs prefer and the \textcolor{red}{red} font refers to the item LLMs struggle.}
  \centering
  \resizebox{0.8\textwidth}{!}{%
  \label{tab:avg-llms-mae}
  \begin{tabular}{lllllllll}
  \toprule
  LLMs    & A4     & A7     & A8     & B4     & B7     & B9     & B10    & B11    \\
  \midrule
  avg-MAE & \textcolor{red}{0.7584} & \textcolor{blue}{0.5557} & \textcolor{red}{0.6492} & \textcolor{blue}{0.6175} & \textcolor{blue}{0.5872} & \textcolor{blue}{0.5748} & \textcolor{red}{0.6697} & \textcolor{red}{0.6912} \\
  \bottomrule
  \end{tabular}%
  }
  \end{table}
  
  \begin{figure}
      \centering
      \includegraphics[width=0.85\linewidth]{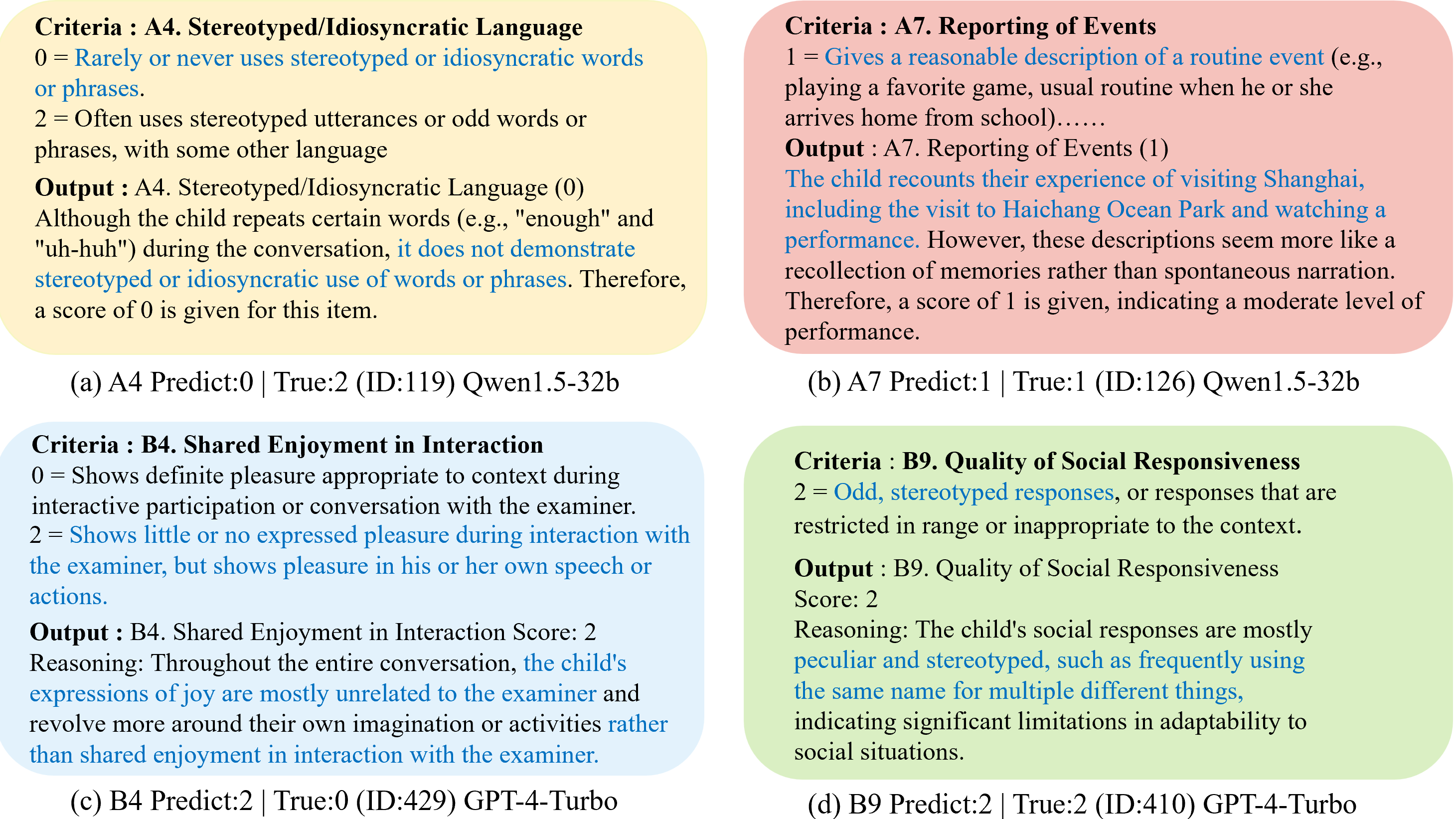}
      \caption{Case study for LLMs'preference Analysis. We provided the scoring items of input and corresponding outputs generated in first stage of our framework. The \textcolor{blue}{blue} font in the figure indicates the basis for scoring.}
      \label{fig:case-analysis}
  \end{figure}
  In this section, we will systematically explain the factors that contribute to the biases of LLMs towards various scoring items. We have calculated the average MAE results for the eight items from all experimental models of its SOTA settings shown in Table ~\ref{tab:avg-llms-mae}. We can observe that LLMs have higher average MAE in the A4, A8, B10, and B11 items, indicating a larger deviation from the clinical diagnoses made by doctors. On the other hand, LLMs show smaller average MAE deviations in the A7, B4, B7, and B9 items, aligning more closely with the results provided by doctors. Based on the ADOS-2 criteria details for the eight items, we can conduct that LLMs demonstrate effectiveness in providing scores for certain items primarily due to two characteristics: Firstly, the ADOS scoring criteria for these items have clear descriptions, and there are evident differences between the scores. Such task descriptions serve as effective prompts to guide LLMs' performance. Secondly, these items are mainly based on objective factors that are less influenced by non-verbal elements.
  
  In Figure ~\ref{fig:case-analysis}, we can observe how LLMs perform scoring: in Figure ~\ref{fig:case-analysis}(a), A4 involves stereotyped and idiosyncratic use of words or phrases, which typically extends beyond just the text and includes aspects of speech and intonation. LLMs may not capture the stereotyped usage solely from the text, leading to an explanation of why they give a score of 0, whereas doctors would consider other factors to cause the bias. In Figure ~\ref{fig:case-analysis}(b), A7 pertains to the reporting of events, and the scoring criteria provide detailed explanations and examples. LLMs, therefore, pay particular attention to the events mentioned in the conversation and provide a relatively reliable score for the overall conversation quality. In Figure ~\ref{fig:case-analysis}(c), B4 involves shared enjoyment in interaction, which is a highly subjective experience that relies on the doctor's expertise and real-time experience. It also encompasses non-verbal aspects. LLMs can only assess the child's expression of joy based on the text and may perceive it as one-sided rather than shared enjoyment with the doctor, leading to a score of 2. However, the doctor's on-site experience could be completely different and might not align with the LLM's assessment. In Figure ~\ref{fig:case-analysis}(d), B9 pertains to the quality of social responsiveness, and its description is concise, impactful, and exhibits significant differences between scores. So LLMs can summarize from the conversation that "the child's social responses are mostly peculiar and stereotyped" and provide an example of "frequently using the same name for multiple different things." Finally, LLMs give a score of 2 based on the evidence of ADOS criteria that "the child demonstrates obvious limitations in adapting to social situations". The explanation is logical, well-supported, and persuasive. Based on the aforementioned analysis, we suggest that the scoring criteria for each assessment item in the ADOS-2 manual should be described in a more detailed and clear manner. It would be highly beneficial to include subjective examples that can help enhance understanding. It is worth mentioning that there is significant variation in the bias of the B4 item across different LLMs. Models such as Qwen1.5 and glm4 scoring more accurately, while models like gpt-4 and Gemini1.5 perform poorly. More detailed discussion of eight item can be found in Appendix ~\ref{sec:what effect}.

  \section{Conclusion}
  In conclusion, this paper explores the potential of LLMs in ASD diagnosis, addressing the limitations of current methods and computer-assisted tools. We propose an evaluation framework called ADOS-Copilot, to enhance LLMs' performance in real-world diagnostic scenarios. Experimental results show that ADOS-Copilot is competitive with doctors' diagnoses, providing evidence-based assessments. Also our findings have significant implications for ASD diagnosis and the broader field of mental health disorders. 
  
  \textbf{Limitations.}
  To facilitate future work, we point out several limitations of our framework. First, it can only be improved by modifying the prompts, unlike doctors who can learn and adapt from their experiences. 
  % The consistency required in institutional scoring can lead to preferences that may not accurately reflect a child's condition, which the framework does not account for. 
  The framework's ability to comprehend and reason with long contexts may not translate to a deep understanding of children's developmental levels, crucial for accurate ASD assessment. The quality of ASR impacts the later processing of diagnostic data, and ASR transcriptions can lose information such as tone and emotion, important for ASD diagnosis. Lastly and most importantly, the framework only uses textual data from ASR transcriptions, but ASD diagnosis is a multi-modal task, involving the assessment of a child's behavior, social interaction, and communication. 
  % The lack of consideration of other modalities limits the framework's comprehensiveness and accuracy.
  
\newpage

\bibliography{paper}

%%%%%%%%%%%%%%%%%%%%%%%%%%%%%%%%%%%%%%%%%%%%%%%%%%%%%%%%%%%%

\newpage

\appendix

% \input{appendix.tex}
% \section{Appendix / supplemental material}

% Optionally include supplemental material (complete proofs, additional experiments and plots) in appendix.
% All such materials \textbf{SHOULD be included in the main submission.} 
% \onecolumn
% {
%     \hypersetup{linkcolor=black}
%     \parskip=0em
%     \renewcommand{\contentsname}{Contents of Appendix}
%     \tableofcontents
%     \addtocontents{toc}{\protect\setcounter{tocdepth}{3}}
% }
\section{Dataset}
\label{sec:dataset}
\subsection{Data collection}

The datasets utilized in this study consist of ADOS-2 (Autism Diagnostic Observation Schedule, Second Edition) assessment processes recorded in a clinical context. After obtaining informed consent from parents or guardians, the complete audio recordings of the ADOS-2 assessment procedures were captured, starting from when the child entered the dedicated assessment room until the child left the room at the end of the evaluation. Any irrelevant data unrelated to the actual assessment, such as the initial and final moments, were removed in subsequent data processing stages. We are committed to protecting the privacy of the children with removing any content that involves children's privacy in data pre-processing phrase.

\subsection{Automatic speech recognition(ASR) and speaker diarization}
Audio transcription is our first step in the whole pipeline.  Our scenario turns out to be a very challenging one for current ASR systems, largely due to children's still developing language abilities and non-standard pronunciation.
We test many commercial services(like iflytek) and open-source ASR models(including
Whisper \cite{radford2022whisper} and Paraformer \cite{gao2022paraformer}), but none of these
give very satisfactory transcriptions, among which large speech models' results are 
more acceptable, just as \cite{du2024speechcolab} indicated. Finally, we choose 
OpenAI's whisper-large-v3 model as our ASR model. 
% \subsection{Speaker Diarization}
Speaker diarization, the process of segmenting audio recordings by speaker labels and aims to answer the question “who spoke when?”,
is also needed since the conversations in ADOS involves multiple speakers. Readers are encouraged to refer to Appendix \ref{sec:asr} for supplementary details on the utilized ASR models.
% \subsubsection{Post-processing transcription using LLMs}
% Inspired by \cite{wang2024diarizationlm},
% we use GPT-4-Tubro to post-process the raw ASR results to enhance the quality of the transcription and add the speakers.(\textbf{P})
\section{ASR samples}
\label{sec:asr}
In this section, we provide an example of a speech transcription text that has undergone preprocessing. Table \ref{tab:transcription} presents a dialogue excerpt between a doctor and a child in the "Friendship, Relationship, and Marriage" task of the ADOS-2-M3 clinical diagnosis. It is observed that there are illogical segments in the dialogue, such as "". However, it is difficult to determine whether these illogical segments stem from the child's diminished social communication abilities associated with Autism Spectrum Disorder (ASD) or if they are a result of inadequate transcription capabilities. Nonetheless, manual transcription methods are extremely expensive, so further exploration of more advanced methods is still necessary.\\
Table \ref{tab:manual transcription} presents a segment of dialogue manually transcribed with speaker diarization. Table \ref{tab:Paraformer transcription} showcases the transcription generated by Paraformer, which notably omits several utterances from the child and contains a considerable number of inaccuracies. Table \ref{tab:whisper transcription} demonstrates the output transcribed by Whisper, a system that, while capable of transcribing the dialogue, fails to identify speakers and also introduces some errors during the transcription process. Lastly, Table \ref{tab:GPT transcription} highlights a transcription that underwent an initial phase of Whisper followed by post-processing with GPT for error correction and speaker labeling. This approach evidences a superior performance compared to Paraformer, effectively illustrating the enhancement in both accuracy and speaker distinction achievable through such a sequential refinement process.

\begin{table}
  \caption{A snip of the transcription}
  \label{tab:transcription}
  \begin{tabular}{l}
    \toprule
    Doctor: I see. So, do you have a boyfriend? \\
    Child: Of course I don't have a boyfriend, but Tongtong is my boyfriend \\
    because he is Yuanyuan, and he is Yuanyuan's cousin.\\
    Doctor: So he is your boyfriend.\\
    Child: He is the boyfriend who came down from the mountain.\\
    Doctor: Oh, I see. How do you know he is your boyfriend?\\
    Child: Because Yuanyuan introduced me to him.\\
    Doctor: Oh, I see. Where do you want to live when you grow up?\\
    Child: You always say not to grow old, so I want to live in my hometown.\\
    Doctor: I mean when you grow up.\\
    Child: When I grow up, oh, I have to go to a university with dorm buildings,\\ 
    so I will live in the dormitory. When I'm old, I will live in a nursing home.\\
    Doctor: Who do you want to live with when you grow up?\\
    Child: I definitely want to live in the dormitory with my classmates.\\
    Doctor: Hmm, have you ever thought about getting married when you grow up?\\
    Child: If I don't want to get married, because the feeling of constantly \\
    having babies after marriage is just too bad.\\
    Doctor: Oh, okay. Do you ever feel lonely?\\
    Child: Sometimes I feel lonely, but then I think about some things I want to do, or things I can't \\
    let go of, and remember some things I've done before, and then I don't feel lonely anymore.\\
    Doctor: Oh, do you think your peers feel lonely?\\
    Child: They don't, they all have friends.\\
    Doctor: So, do you feel lonely?\\
    Child: I don't feel lonely.\\
    \bottomrule
  \end{tabular}
\end{table}

\begin{table}
  \caption{A snip of the transcription by manual annotation}
  \label{tab:manual transcription}
  \centering
  \begin{tabular}{l}
    \toprule
    Doctor: Have you finished putting it together then?\\
    Child: Yes.\\
    Doctor: I think you've done a fantastic job!\\
    Doctor: Look, there's an airplane.\\
    Child: Hmm, why is there a poison symbol here?\\
    Doctor: Oh, this is the poison sign, do you recognize it?\\
    Doctor: Ah, because this is an eco-friendly bag.\\
    Doctor: And actually, this isn't poison.\\
    Doctor: This is a recycling symbol.\\
    Doctor: Where did you see this poison symbol?\\
    Child: I saw it on a barrel.\\
    Doctor: On a barrel?\\
    Child: That barrel was a car, actually.\\
    Child: It was on the cement mixer truck where I saw the poison symbol.\\
    Doctor: On the cement mixer truck?\\
    Doctor: There might have been hazardous materials, right?\\
    Child: Missiles.\\
    Doctor: Would there be missiles on a mixer truck?\\
    Child: No, but the mixed cement could be toxic.\\
    Doctor: The mixed cement is toxic?\\
    Doctor: How do you know that?\\
    Doctor: I don't seem to know.\\
    Child: Because it's all marked there.\\
    \bottomrule
  \end{tabular}
\end{table}

\begin{table}
  \caption{A snip of the transcription by Paraformer\cite{gao2022paraformer}}
  \label{tab:Paraformer transcription}
  \centering
  \begin{tabular}{l}
    \toprule
    Doctor: Did you listen carefully?\\
    Doctor: I think your assembly is wonderful.\\
    Doctor: Look, there’s an airplane.\\
    Doctor: This is a toxic symbol, do you recognize it?\\
    Doctor: It's an eco-bag. This is not about toxicity but a recycling symbol.\\
    Doctor: Where did you see the toxic symbol?\\
    Doctor: On a barrel.\\
    Doctor: On a cement mixer, perhaps carrying dangerous goods.\\
    Child: Missiles.\\
    Doctor: Missiles.\\
    Doctor: Missiles on a mixer truck?\\
    Doctor: No, the produced concrete is poisonous.\\
    Doctor: How do you know?\\
    Doctor: I don't know where exactly,\\
    Doctor: But it's all indicated there.\\
    \bottomrule
  \end{tabular}
\end{table}

\begin{table}
  \caption{A snip of the transcription by Whisper\cite{radford2022whisper}}
  \label{tab:whisper transcription}
  \centering
  \begin{tabular}{l}
    \toprule
    Are you done assembling it then?\\
    I think you've done a great job.\\
    I see an airplane.\\
    Hmm, why is there a toxic substance here?\\
    Oh, this is the toxic sign, do you know it?\\
    Ah, this is an environmentally friendly bag.\\
    And this actually isn't toxic.\\
    This is a recycling symbol.\\
    Where did you spot this toxic symbol?\\
    Saw it on a bucket.\\
    On a bucket.\\
    That bucket happened to be a car,\\
    Specifically, saw that toxic symbol on the cement mixer.\\
    On the cement mixer.\\
    Could be dangerous goods,\\
    Right?\\
    Missiles.\\
    Would there be missiles on a mixer?\\
    No, but the mixed concrete would be poisonous.\\
    Mixed concrete is poisonous?\\
    How do you know?\\
    I don't seem to know.\\
    Because it's all labeled there.\\
    \bottomrule
  \end{tabular}
\end{table}

\begin{table}
  \caption{A snip of the transcription by Whisper+GPT refinement \cite{radford2022whisper}}
  \label{tab:GPT transcription}
  \centering
  \begin{tabular}{l}
    \toprule
    Doctor: Have you completed the assembly? I must say, your work is impressive. \\I spotted an airplane there.\\
    Child: Why is there a toxic symbol here, though?\\
    Doctor: Ah, you recognize the toxic sign? That's interesting.\\
    Child: Oh, because this is an eco-bag, and actually, this isn’t toxic—it's a \\recycling symbol.\\
    Doctor: Where did you first see the toxic symbol?\\
    Child: On a barrel, which was a car—a cement mixer with that toxic symbol.\\
    Doctor: On a cement mixer, possibly carrying hazardous materials, yes?\\
    Child: Missiles.\\
    Doctor: Missiles on a mixer?\\
    Child: Not missiles, but the mixed concrete might be toxic.\\
    Doctor: Toxic concrete? How did you learn that?\\
    Child: I'm not sure, but everything is clearly labeled.\\
    \bottomrule
  \end{tabular}
\end{table}

\section{Introduction of ADOS-2}
\label{sec:ados-2}
% \usepackage{booktabs}
\begin{table}[]
\caption{Statistics of our dataset\,(28 samples in total)}
\label{tab:stat}
\centering
\begin{tabular}{@{}ll@{}}
\toprule
Item       & statistics                                          \\ \midrule
gender     & boy:20(71.4\%)  girl:8(28.6\%)                      \\
age(months)        & 93.6±28.2  (min=57,max=173)                   \\
ADOS score & 7.25±4.56 (min=1,max=17)                             \\
2-cls      & TD: 12(42.9\%)  ASD: 16(57.1\%)                     \\
3-cls      & TD: 12(42.9\%)  ASD: 4(14.3\%)  Austim: 12(42.9\%) \\ \bottomrule
\end{tabular}
\end{table}
\subsection{Procedure}
The ADOS-2 consists of a series of structured and semi-structured tasks, typically taking 40-60 minutes to administer. During this time, the examiner provides opportunities for the subject to demonstrate social and communication behaviors relevant to an autism diagnosis. Activities are selected from the module that matches the subject's developmental and language level. Module 3, which includes 14 activities (see \ref{tab:ADOS-2 M3}), is suitable for verbally fluent children or young adolescents.

% \usepackage{booktabs}
% \usepackage{graphicx}
\begin{table}[]
\centering
\caption{ADOS-2 Module 3 Activities}
\label{tab:ADOS-2 M3}
\resizebox{\textwidth}{!}{%
\begin{tabular}{@{}ll@{}}
\toprule
Activities &
  Summary \\ \midrule
Construction Task &
  Provide a warm-up activity to observe the interactive behavior of the   subjects. \\
Make Believe Play &
  Observe the extent to which the subjects use toys and dolls creatively in   an unstructured task. \\
Joint Interactive Play &
  Assess the extent and quality of the subjects' coordination of behavior   and emotional expression with the assessor. \\
Demonstration Task &
  Assess the ability of the subject to use a set of actions accompanied by   gestures or language, and report a familiar event. \\
Description of Picture &
  Obtain an example of the subject's spontaneous language and   communication, and understand what can attract his interest. \\
Telling a Story from a   Book &
  Assess the ability to describe a sequential story from a picture and comment on social relationships and emotional expression. \\
Cartoons &
  Observe how the subject narrates a   story, uses gestures to present events, \\
Conversation and   Reporting &
  Assess the subject's ability to engage in back-and-forth conversation and   describe an event or situation without visual cues. \\
Emotions &
  The subject details two emotions, their triggers, and related personal experiences. \\
Social Difficulties and   Annoyance &
  Assess the subject's insight into their own social difficulties and their   sense of responsibility for their own behavior. \\
Break &
  Provide the subject with a break from the structured social demands of   the assessment and observe their behavior in a less structured setting. \\
Friends and Marriage &
  The subject details one or more types of relationships and shares their   views on friendship, family or a long-term partnership. \\
Loneliness &
  Assess the subject's insight into   their own social situations and their ability to describe their emotional   responses. \\
Creating a Story &
  The subject uses objects creatively to tell a new story. \\ \bottomrule
\end{tabular}%
}
\end{table}

\subsection{Scoring}

The scoring of ADOS does not strictly follow a one-to-one task scoring method. Instead, it evaluates the subject's overall performance throughout the entire assessment process. This is why we have introduced LLMs to handle the complete ADOS-2 assessment process, rather than using only a portion of its content. Here, we will only discuss the specific scoring criteria actually used in this experiment.

\subsubsection{A4}
This item's assessment includes delayed echolalia or other highly repetitive speech patterns with consistent intonation. These words or phrases can be meaningful and can be somewhat applicable to conversation. The focus of this project is the stereotyped or characteristic nature of the phrasing.

0 = Rarely or never uses stereotyped or idiosyncratic words or phrases.

1 = Uses vocabulary or phrases that are more repetitive or formulaic than most people at the same level of expressive language ability, but not noticeably strange, or occasionally uses strange uses of stereotyped discourse or phrases, with substantial spontaneous and flexible language.

2 = Frequently uses stereotyped discourse or strange words or phrases, along with some other language.

3 = Almost exclusively uses strange or stereotyped language. And rarely uses non-stereotyped natural language.

\subsubsection{A7}
This item focuses on the participant's ability to spontaneously select an event or respond to a general question from the assessor and describe it in a clear and understandable way, without requiring specific prompting. This involves a sequential description of events that occurred outside of the immediate environment.

0 = Reports a specific, real, and unconventional event in a non-stereotyped manner (e.g., a vacation, a trip, a shopping spree). The participant provides this description without being directly asked, but may need an initial general question to get started.

1 = Provides a well-structured description of a routine event (e.g., what they did on the way home from school, playing their favorite game). This is not related to their profession or hobbies and is likely a truthful account. The participant offers this description without prompting, but may initially need to be asked to elaborate on the event. Descriptions from the "Demonstration Task" are also included here.

2 = Provides a description of a regular or unconventional event, but relies on specific prompting from the assessor, or describes an event that seems unlikely to be true.

3 = Response to specific prompting is inconsistent or insufficient.

\subsubsection{A8}
This summary item evaluates the participant's ability to engage in reciprocal conversation. The rating should encompass all conversation opportunities, not just the most successful ones.

0 = Fluent conversation builds upon the assessor's contributions. The participant primarily delivers responses that address the assessor's prompts and adds their own thoughts or questions based on the ongoing dialogue. This allows for a back-and-forth exchange with at least four elements: assessor initiates, participant comments, assessor responds, and participant responds.

1 = The conversation includes some elaboration by the participant on their responses or provides cues to guide the assessor's direction. However, this elaboration may be limited in quantity or flexibility compared to what's expected at the participant's expressive language level.

2 = The participant initiates minimal reciprocal conversation and may prioritize their own train of thought over actively engaging. While some spontaneous information or comments might be present, there's a weak sense of true back-and-forth exchange.

3 = The participant demonstrates very little spontaneous communicative language (though there may be a significant amount of repetitive or non-communicative language). This rating applies to participants with limitations who struggle to respond meaningfully to the assessor's conversation openers.

\subsubsection{B4}
This item assesses the participant's ability to express pleasure to the assessor, beyond simply interacting or responding.

0 = Exhibits pleasure in at least one task or conversation topic during engagement or interaction with the assessor.

1 = Exhibits context-appropriate pleasure during interactions with the assessor or shows clear pleasure in one interaction.

2 = Shows little or no expressed pleasure in interactions with the assessor, but may show pleasure in their own speech or behavior or in non-interactive components of the ADOS-2 materials or activities.

3 = Expresses very little or no pleasure throughout the ADOS-2 assessment.

\subsubsection{B7}
This summary item focuses on the quality of the participant's attempts to initiate social interaction with the assessor, rather than the frequency of such attempts.

0 = Effectively uses nonverbal and verbal means to communicate clear social overtures to the assessor. These overtures must be appropriate for the current context.

1 = Some social overtures are slightly unusual in nature. Initiations may be limited to personal needs or related to the participant's own interests, but in some cases, may draw the assessor in.

2 = Inappropriate overtures; many overtures lack the ability to blend into the context and/or have social quality. This includes the focus of the participant's overtures, with few attempts to draw the assessor in.

3 = No social overtures of any kind.

\subsubsection{B9}
Another summary item that evaluates the participant's social responses.

0 = Exhibits a range of appropriate responses that vary depending on the current social context and pressure.

1 = Responds to most social cues but with some limitations, inappropriateness, inconsistency, or consistent negativity.

2 = Strange, stereotyped, or limited-range responses, or responses that are not contextually appropriate.

3 = Little or no response to the assessor's attempts to engage the participant.

\subsubsection{B10}
This item focuses on the frequency of interaction using any communication mode during the ADOS-2 assessment. Frequency here is defined by the number of occurrences and distribution across a range of contexts. The rating for this summary project should describe all aspects of nonverbal and verbal/vocal behavior (which do not need to be coordinated), but there must be at least a brief interaction with the assessor (not others who may be present in the ADOS-2 assessment room).

0 = Extensive use of verbal or nonverbal behavior (regardless of the level of proficiency) for social interaction (i.e., chatting, commenting, making remarks, or nonverbal behavior that appears to be reciprocal).

1 = Some reciprocal social interaction (as described in the 0 rating above), but with a reduced frequency or quantity or the number of contexts in which such interactions occur (regardless of the amount of non-social talk).

2 = Most communication is either object-oriented (i.e., asking about things), answering questions, echolalic, or has a specific premise, with little or no social interaction/back-and-forth talk.

3 = Very little or no interaction with the assessor.

\subsubsection{B11}
This is another summary item that reflects the overall judgment of the rapport that the assessor establishes with the participant during the ADOS-2 assessment. Particular consideration is given to the extent to which the assessor has to change their own behavior in order to successfully maintain the interaction.

0 = Comfortable interaction between the participant and the assessor, appropriate for the context of the ADOS-2 assessment.

1 = Interaction is sometimes comfortable but not sustained (e.g., may feel awkward or constrained at times, or the participant's behavior may seem mechanical or slightly inappropriate).

2 = One-sided or unusual interactions lead to a conversation that is persistently mildly uncomfortable, or the conversation would be difficult if the assessor did not repeatedly change this outside of the standard activities in the ADOS-2 assessment.

3 = The participant shows very little attention to the assessor, or the conversation is noticeably uncomfortable for an extended period of time.

\subsection{Classification}
In clinical practice, children's scores are limited to 0, 1, and 2 (with a score of 3 being merged into 2). For Module 3, the final ADOS-2 M3 score consists of social affect(SA), communication (3 items, 0-6 points) and reciprocal social interaction (7 items, 0-14 points), and restricted and repetitive behaviors(RRB) (4 items, 0-8 points). This results in a total of 14 items and a score range of 0-28 points. The ADOS-2 uses a cutoff method for diagnosing ASD in children. Specifically, for children assessed with M3, a total score of 0-6 indicates a non-spectrum disorder, 7-8 indicates a spectrum disorder, and 9 or above indicates Autism. The scoring items, standards, and cutoff algorithms vary between modules according to the child's developmental level. 

\section{More details for prompts using by ADOS-Copilot}

\subsection{Concise vs standard criteria}
\label{sec:concise}
\begin{figure}
    \centering
    \includegraphics[width=1\linewidth]{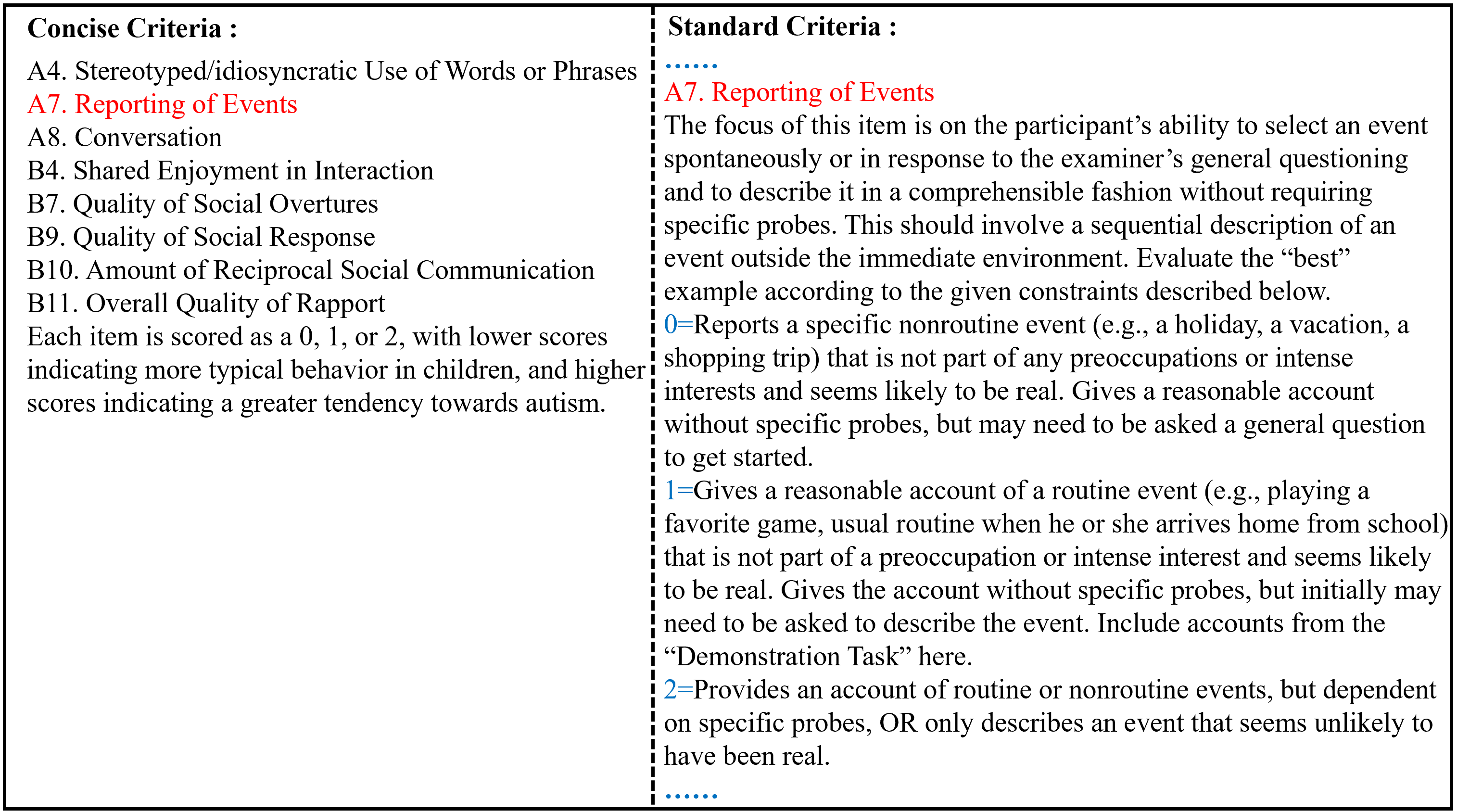}
    \caption{Concise vs Standard Criteria of ADOS-2. In the ablation experiment, we compared the effects of the concise criteria and standard criteria. This Figure provides specific examples of both to better understand the differences in their effects.}
    \label{fig:criteria}
\end{figure}

Figure \ref{fig:criteria} shows a comparison of the concise and standard(\textit{C}) prompts used in the ablation experiment. It is evident that the concise criteria only include the item name and an overall scoring description, whereas the standard criteria provide a detailed item description and corresponding criteria justifications for each score.

\subsection{Details for prompt of in-context enhancement}
\begin{figure}
    \centering
    \includegraphics[width=0.9\linewidth]{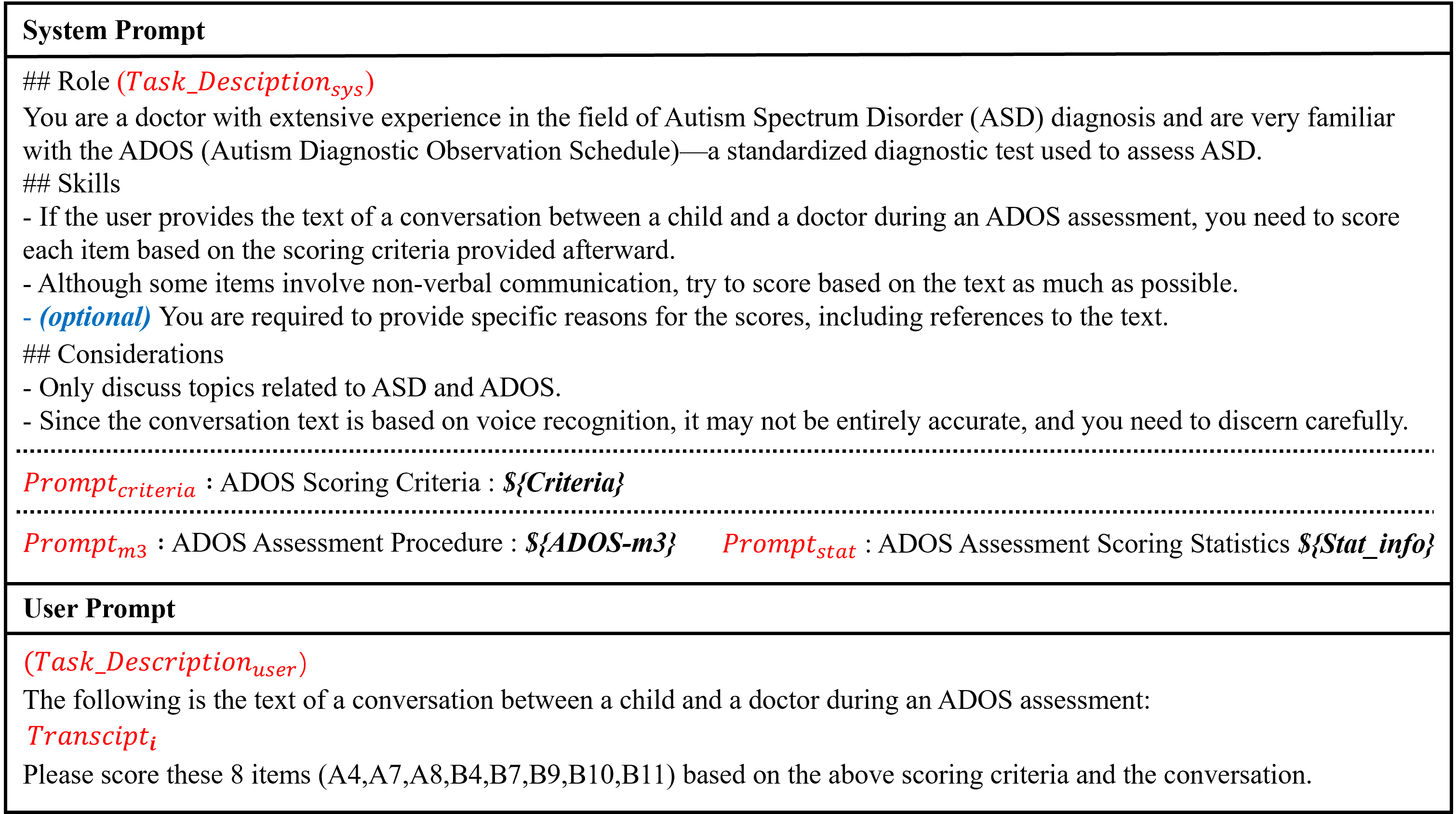}
    \caption{In-context Enhancement prompt for ADOS-Copilot. Where \(Prompt_{criteria}\) refers to the prompt of clinical ADOS criteria, \(Prompt_{m3}\) refers to the prompt of clinical ADOS-2 Module 3 procedures, \(Prompt_{stat}\) refers to the prior information of the ASD and TD children, \(Transcript_i\) refers to the pro-processing dialogue texts between doctor and child.}
    \label{fig:base prompt}
\end{figure}

As shown in Figure \ref{fig:base prompt}, we can see more detail of our in-context enhancement prompt.
\label{sec:in-context}
% By leveraging the contextual learning ability\cite{brown2020language}, LLMs are capable of completing the various task without updating parameters. Therefore, we introduce the criteria and procedures of the ADOS-2 as well as the prior statistical infomation as in-context enhancement prompts. For criteria, some researches have shown that LLMs tend to make more errors when confronted with more complex conversational contexts\cite{bhattacharya2023context,li2023compressing,xu2024mental}. Therefore, we introduce more detailed criteria to guide LLMs in properly activating existing clinical knowledge to better understand this complex task. Inspired by \cite{qin2023read}, we incorporate the scoring criteria of the eight items used in clinical ASD diagnosis into the system prompt.
% For procedures, we consider that the evaluation of ADOS-2 consists of a series of consecutive scenario-based interactions. However, it is challenging to segment the individual scenes during the process of speech recognition. In order to facilitate the model's understanding of the evaluation scenarios in ADOS-2, we include excerpts from the assessment process manual used as a reference in hospitals as part of the prompt. For prior information, to better align with the assessment results of doctors and mitigate potential biases inherent in LLMs, we introduce prior statistical information into the prompt. The statistics are collected by hospital, including the mean of each item and the proportion of ASD and TD children. 

\subsection{Prompt of second-stage for interpretability augmentation}
\label{sec:second stage prompt}
\begin{figure}
    \centering
    \includegraphics[width=1\linewidth]{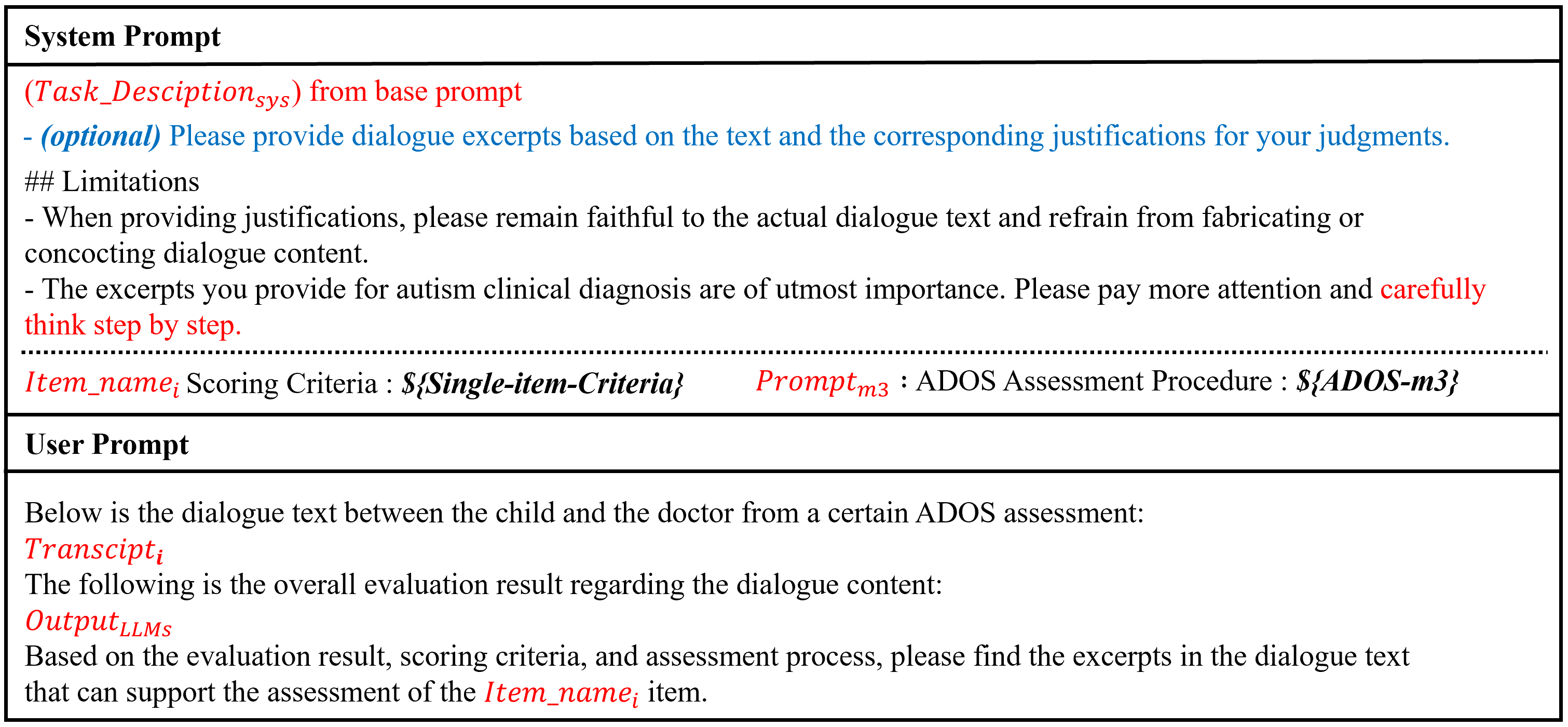}
    \caption{Prompt of second-stage for Interpretability Augmentation}
    \label{fig:two-stage-prompt}
\end{figure}

As shown in Figure \ref{fig:two-stage-prompt}, We replaced the \textcolor{blue}{-(optional)} content from our prompt and enhanced the interpretability separately for each individual scoring item. We extract the single-item criteria by separating the items in ADOS Scoring Criteria. Then, we incorporated the scores and justifications generated by LLMs called \(Output_{LLMs}\) in the first stage into the user prompt.

% \begin{table}
%   \caption{An item of ADOS-2 scoring criteria}
%   \label{tab:scoring criteria}
%   \begin{tabular}{l}
%     \toprule
%     \textcolor{red}{A7.Reporting of Events} \\
% The focus of this item is on the participant’s ability to select an event spontaneously or in \\response to the examiner’s general questioning and to describe it in a comprehensible fashion \\without requiring specific probes. This should involve a sequential description of an event \\outside the immediate environment. Code the “best” example, given the rating constraints described \\below with regard to preoccupations and probes.\\
% \textcolor{blue}{0 = }Reports a specific nonroutine event (e.g., a holiday, a vacation, a shopping trip) that is not \\part of any preoccupations or intense interests and seems likely to be real. Gives a reasonable \\account without specific probes, but may need to be asked a general question to get started.\\
% \textcolor{blue}{1 = }Gives a reasonable account of a routine event (e.g., playing a favorite game, usual routine \\when he or she arrives home from school) that is not part of a preoccupation or intense interest \\and seems likely to be real. Gives the account without specific probes, but initially may need to \\be asked to describe the event. Include accounts from the “Demonstration Task” here.\\
% \textcolor{blue}{2 = }Provides an account of routine or nonroutine events, but dependent on specific probes, OR only \\describes an event that seems unlikely to have been real.\\
%     \bottomrule
%   \end{tabular}
% \end{table}

\begin{table}
  \caption{An excerpt from the scoring description of the ADOS-2 procedures}
  \begin{tabular}{l}
    \toprule
    2.	Make-Believe Play
\\Purpose: This activity provides the opportunity to observe the participant’s creative or \\imaginative use of miniature play objects in an unstructured task.
\\Materials: Contents of Bag 3—two male action figures and a female action figure, with one “prop” \\for each; miniature hairbrush; two small tools; toy dinosaur. Contents of Bag 2—small spoons and \\plates, several pieces of miniature food, small teapot/pitcher/measuring cup, miniature book, toy \\car, toy rocket, small ball, hologram disk, and two pieces of “junk” (a small piece of cloth and a \\small “jewelry” box). If necessary, other materials from Module 2 may be added (i.e., from Bag 1), \\but because they require somewhat less creativity, the difference in contexts from the standard \\presentation should be taken into account when coding.
\\Instructions:
\\-Lay out the materials from Bag 3. Introduce the action figures with descriptions appropriate to \\their appearance (e.g., “Here are a princess, a wrestler, and a soldier, and their pet dinosaur”).
\\-Lay out the materials from Bag 2, saying, “Here are some of their things. Can you play with these \\for a while?”
\\-Observe the participant’s behavior. If the participant does nothing or seems uncomfortable, then \\after a few moments, say, “I’ll play with these.”\\
    \bottomrule
  \end{tabular}
\end{table}

\section{Feature and paradigm design of rule-based model}
\label{sec:paradigm design}
\textbf{Feature design}\\
In clinical practice, clinicians need to score children's performance based on ADOS-2 annotations. However, current research has not adequately addressed the objective and quantitative assessment of these issues\cite{cheng2023computer}. To identify the risk of ASD in children by analyzing their language patterns, we propose a set of evaluation features based on ADOS-2 and clinical practices inspired by \cite{molloy2011use}. These features quantify children's language communication and reciprocal Social interaction. Predefined scoring rules then assign evaluation scores to child participants. The evaluation features are introduced as follows:

\textbf{Child immediate echolalia rate}
This feature assesses the severity of immediate echolalia in children. We utilize the levenshtein distance\cite{yujian2007normalized} to determine whether the child's speech mimics the doctor's utterances.

\textbf{Conversation alternating rate}
This feature evaluates the flow of conversation between the child and the doctor. It reflects the smoothness of the conversation based on the frequency of turn-taking between the child and the doctor in the dialogue text.

\textbf{Child participant rate}
This feature assesses the child's level of participation in the conversation. It measures the proportion of the child's speech in the dialogue text to determine the child's engagement in the conversation.

\textbf{Child enjoyment rate}
This feature evaluates the child's ability to express enjoyment. By conducting sentiment analysis\cite{gronemeyer202320} on the child's speech, we assess the degree of enjoyment during the evaluation based on the proportion of positive utterances by the child.

\textbf{Child passive rate}
This feature assesses the level of passivity in the child. Through sentiment analysis of the child's speech, we evaluate the degree of child's negative emotions during the assessment based on the proportion of negative utterances by the child.

\textbf{Child social suggestion rate}
This feature evaluates the frequency with which the child offers social suggestions. We determine the child's tendency to provide social suggestions based on the proportion of suggestions within the total number of their utterances.

\textbf{Child social response rate}
This feature assesses the frequency of the child's social responses. We measure the child's active social responding by calculating the proportion of the child's responses to the doctor's questions relative to the total number of questions asked.

\begin{figure*}
  \centering
    \subfigure[A4]{\includegraphics[width=0.24\textwidth]{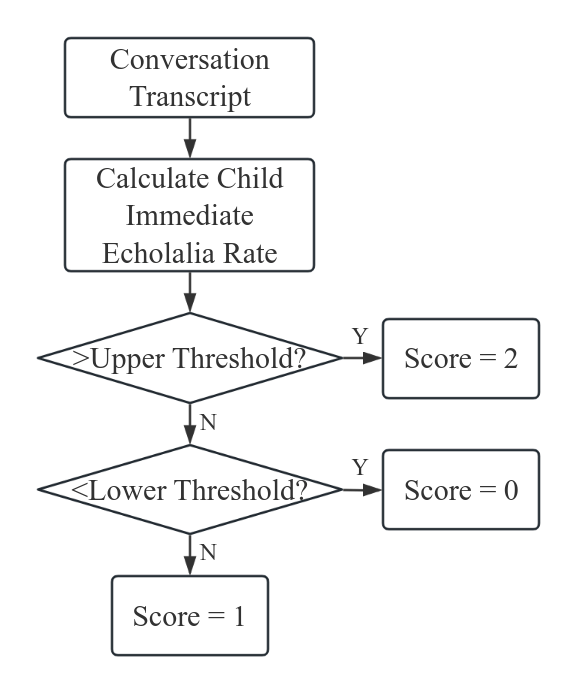}} 
	\subfigure[A7]{\includegraphics[width=0.24\textwidth]{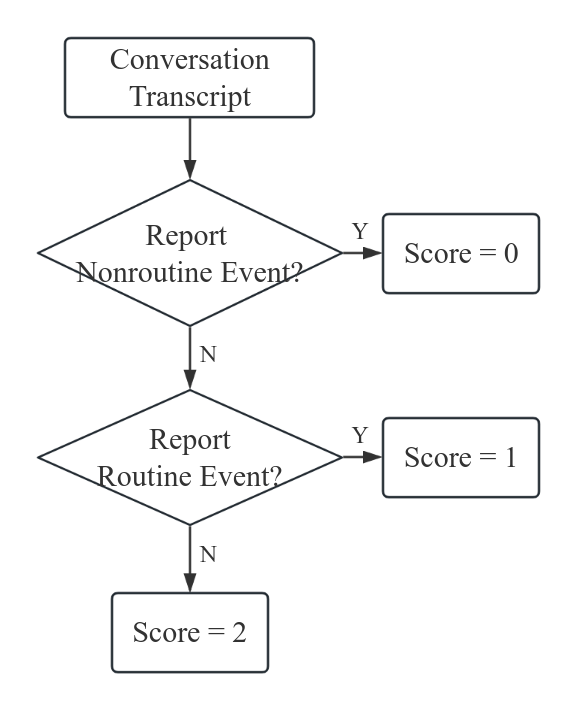}} 
    \subfigure[A8]{\includegraphics[width=0.24\textwidth]{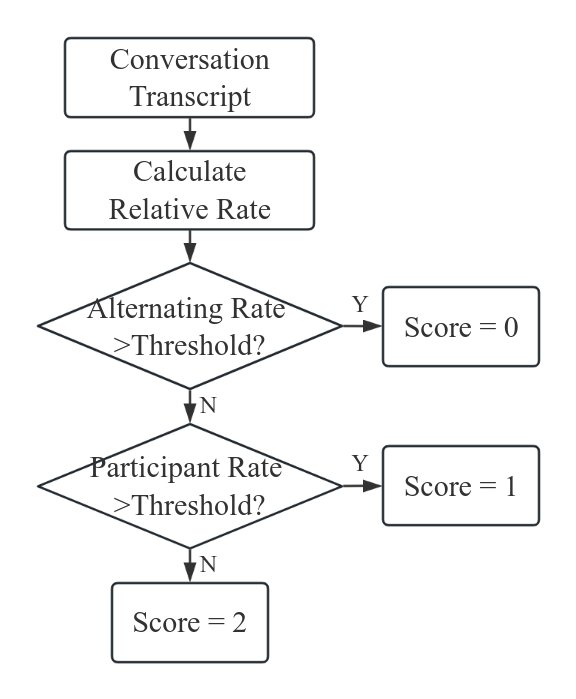}} 
	\subfigure[B4]{\includegraphics[width=0.24\textwidth]{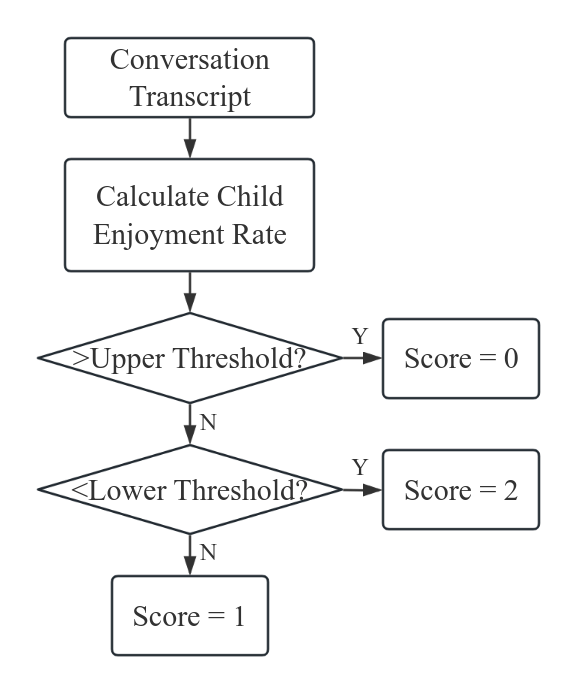}} \\
    \subfigure[B7]{\includegraphics[width=0.24\textwidth]{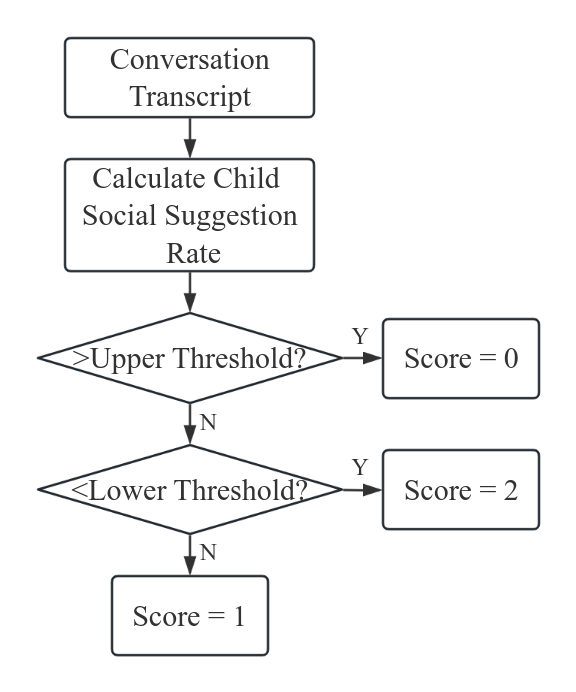}} 
	\subfigure[B9]{\includegraphics[width=0.24\textwidth]{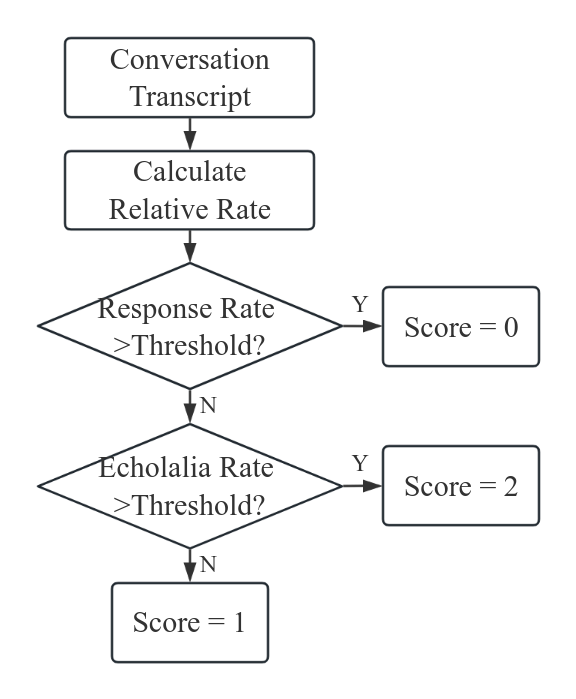}} 
    \subfigure[B10]{\includegraphics[width=0.24\textwidth]{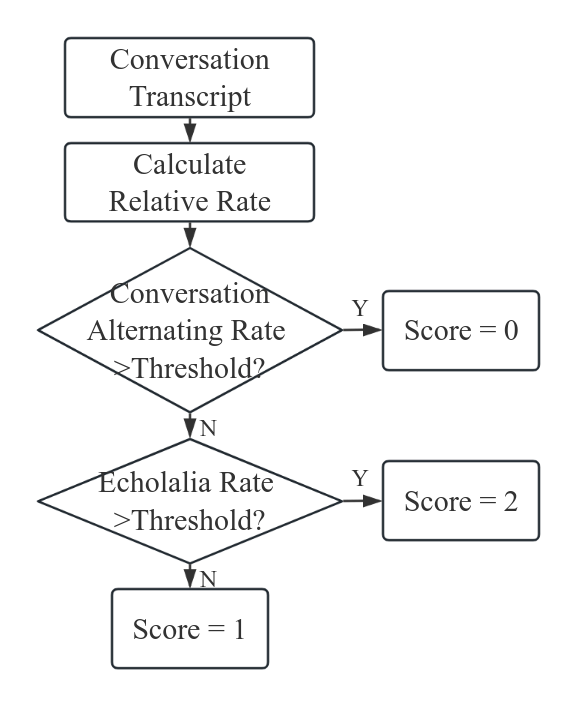}} 
	\subfigure[B11]{\includegraphics[width=0.24\textwidth]{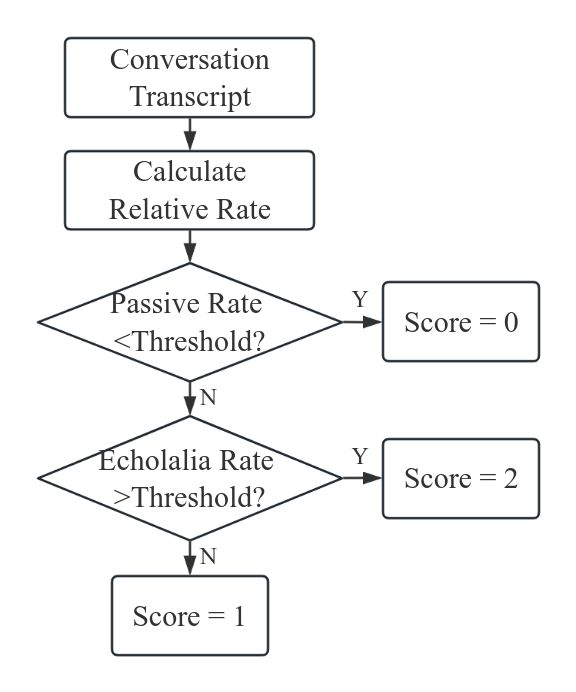}} \\
  \caption{Paradigm design flow chart. Please refer to Appendix \ref{sec:ados-2} for the scoring criteria of specific items}
	\label{fig:paradigm design}
	\vspace{0.2in}
\end{figure*}

\textbf{Paradigm design}\\
Figure \ref{fig:paradigm design} illustrates the scoring rules we devised for various items, informed by the ADOS-2 documentation and clinical practice. To ascertain the threshold parameters for these rules, we employed a stratified two-fold cross-validation approach coupled with grid search. The mean Mean Absolute Error (MAE) on the validation sets was adopted as the primary evaluation metric, ensuring a robust and comprehensive assessment of our model's performance across different subsets of the data. This strategy not only validates the scoring rules' Generality but also optimizes their effectiveness by fine-tuning the parameters against real-world variability captured within the validation folds.

\section{More details of our proposed framework via case study}
\label{sec:case study}
\begin{figure}
    \centering
    \includegraphics[width=1\linewidth]{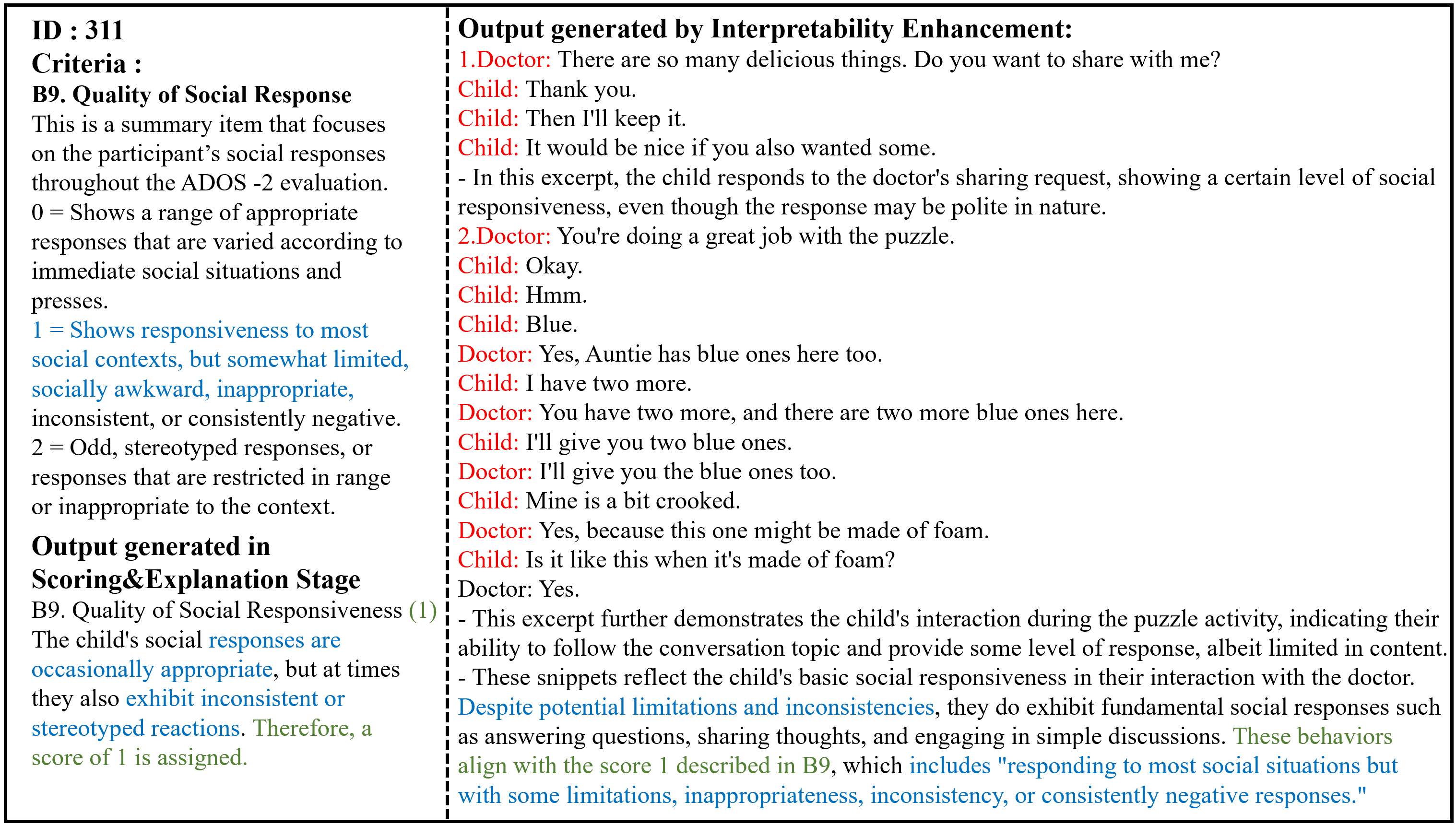}
    \caption{Case Study for B9 Scoring and Explanation. The \textcolor{blue}{blue} part refers to the references to the ADOS-2 scoring criteria, the \textcolor{green}{green} part signifies scoring consistency, and the \textcolor{red}{red} part indicates several original dialogue segments that support the scoring decisions made by our framework.}
    \label{fig:appendix-two-stage}
\end{figure}

Figure \ref{fig:appendix-two-stage} illustrates an example of our proposed framework based on Qwen1.5-32b for assessing B9 (Quality of Social Responsiveness). In the Scoring\&Explanation stage, ADOS-Copilot determines that the child's social responses still exhibit inconsistency and stereotyped reactions, aligning with the description for the score of 1. In the Interpretability Augmentation stage, two excerpts are extracted from the original dialogue to provide evidence for the reasoning behind the score. These excerpts intuitively demonstrate that the child does respond to the doctor's interaction, but the responses are brief and not entirely appropriate, further validating the assessment in the first stage and reinforcing the objectivity and rationality of assigning a score of 1. This showcases the superiority of our proposed framework in scoring and explanation.

% \section{More cases of explanation augmentation stage}
\section{More analysis of experiment results}
\label{sec:more-results}
\subsection{Detailed results of rule-based and fusion model}

\begin{table}[]
\caption{Results of different fusion strategy}
\label{tab:fusion-multiversion-result}
\resizebox{\textwidth}{!}{%
\begin{tabular}{@{}llllllllllllllll@{}}
\toprule
\multicolumn{1}{c}{\textbf{model}} &
  \multicolumn{1}{c}{\textbf{A4}} &
  \multicolumn{1}{c}{\textbf{A7}} &
  \multicolumn{1}{c}{\textbf{A8}} &
  \multicolumn{1}{c}{\textbf{B4}} &
  \multicolumn{1}{c}{\textbf{B7}} &
  \multicolumn{1}{c}{\textbf{B9}} &
  \multicolumn{1}{c}{\textbf{B10}} &
  \multicolumn{1}{c}{\textbf{B11}} &
  \multicolumn{1}{c}{\textbf{avg}} &
  \multicolumn{1}{c}{\textbf{2-acc}} &
  \multicolumn{1}{c}{\textbf{2-precision}} &
  \multicolumn{1}{c}{\textbf{2-f1}} &
  \multicolumn{1}{c}{\textbf{3-acc}} &
  \multicolumn{1}{c}{\textbf{3-precision}} &
  \multicolumn{1}{c}{\textbf{3-f1}} \\ \midrule

Fusion-v1 & \textbf{0.5714} & \textbf{0.4643} & 0.5000 & 0.2500 & 0.3214 & \textbf{0.6071} & \textbf{0.6071} & \textbf{0.6786}  & 0.4805 & 0.6786 & 0.7278 & 0.6498 & 0.5357 & 0.6748 & 0.5610 \\
Fusion-v2 & 0.6859 & 0.5152 & 0.4825 & 0.1894 & 0.3037 & 0.5398 & 0.6235 & 0.6947 & 0.4642 & \textbf{0.8214} & \textbf{0.8661} & 0.8129 & \textbf{0.7500} & 0.8626 & \textbf{0.7837} \\
Fusion-v3 & 0.6755 & 0.5125 & 0.4828 & 0.1912 & 0.3037 & 0.5439 & 0.6220 & 0.6930 & \textbf{0.4633} & \textbf{0.8214} & \textbf{0.8661} & 0.8129 & \textbf{0.7500} & 0.8626 & \textbf{0.7837} \\
Fusion-v4 & 0.6885 & 0.5165 & \textbf{0.4823} & 0.1877 & \textbf{0.3036} & \textbf{0.5383} & 0.6240 & 0.6951 & 0.4643 & \textbf{0.8214} & \textbf{0.8661} & 0.8129 & \textbf{0.7500} & 0.8626 & \textbf{0.7837} \\ 
\bottomrule
Rule     & 0.6938 & 0.5137 & 0.5190 & \textbf{0.1178} & 0.3245 & 0.6809 & 0.7196 & 0.8252 & 0.4982 & 0.7903 & 0.8271 & 0.7887 & 0.6854 & 0.7426 & 0.6833 \\
LLM      & 0.8214 & \textbf{0.4643} & 0.5000 & 0.1250 & 0.3214 & 0.6071 & \textbf{0.6071} & \textbf{0.6786} & 0.4756 & \textbf{0.8214} & 0.8345 & \textbf{0.8179} & 0.7143 & \textbf{0.8627} & 0.7451 \\
\bottomrule
\end{tabular}%
}
\end{table}

The presented experimental data in Table \ref{tab:fusion-multiversion-result} offer insights into the performance of various fusion strategies (Fusion-v1 through Fusion-v4) across multiple items (A4 to B11) and aggregated performance metrics, contrasting these with standalone rule-based and LLM models. The analysis underscores the impact of different weighting schemes on fusion efficacy.

The rule-based model excels in particular tasks (e.g., B1, B2), as reflected in high accuracy, highlighting its domain-specific precision. However, its modest overall average accuracy points to limitations in generalizability beyond these niches. The LLM demonstrates exceptional performance in certain categories (like A4), yet lags in overall average accuracy and other comprehensive metrics when compared to fused models, indicating a trade-off between specialized accuracy in isolated instances and broader consistency.

Fusion-v1 employs a rudimentary approach, assigning a coefficient of 1 to the model with lower MAE and 0 to the other, effectively silencing one model in the decision-making process. While this straightforward method shows promise in specific categories (such as A4 and A7), its overall average accuracy (avg) and composite metrics lag behind subsequent fusion techniques, suggesting that linear weighting may underutilize the complementary strengths of the two models.

Fusion-v2, v3, and v4 refine the weighting strategy by applying the inverse, square of the inverse, and softmax transformations of MAE values, respectively. These advanced schemes yield improvements across most indicators, with Fusion-v2 and v3 notably excelling in average accuracy, 2-acc, and 2-precision, illustrating the potency of nuanced weight allocation in harnessing the advantages of both models. Fusion-v4, despite adopting a more sophisticated softmax weighting, delivers comparable results, implying that simpler inversions (as in v2 and v3) might suffice for optimal performance in this context.

\subsection{Effect of language}
\begin{figure}
    \centering
    \includegraphics[width=0.7\linewidth]{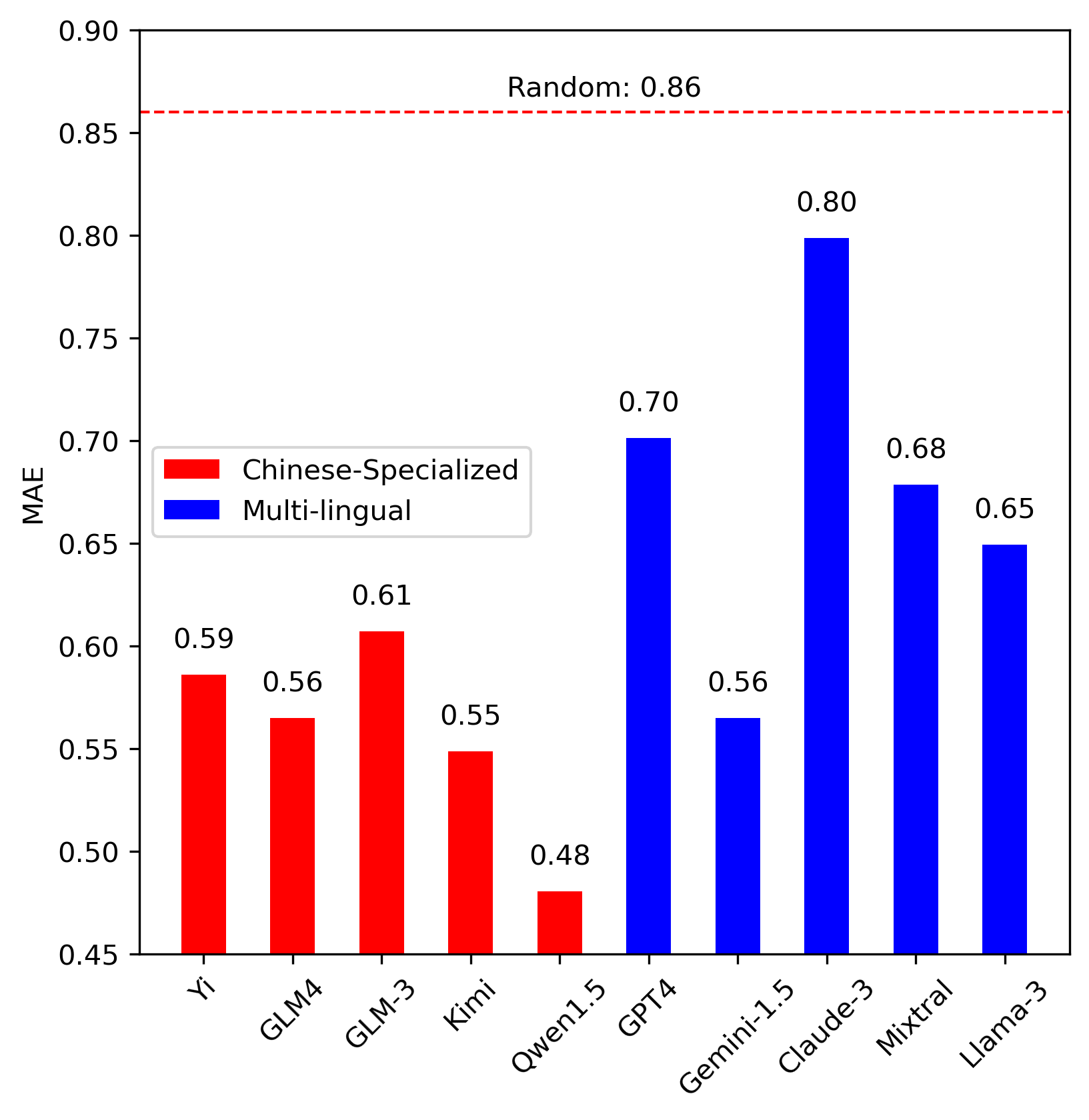}
    \caption{Effect of language. We compare the average MAE of 8 items of 5 Chinese-specialized and 5 multi-lingual famous LLMs.}
    \label{fig:lang}
\end{figure}

Apparently in Figure \ref{fig:lang}, LLMs that specialzes in Chinese perform better than multi-lingual ones in general. This phenomenon is expected, as many papers suggest \cite{sun2024benchmarking}. Even GPT-4 undergo much disadvantage in our task.  We believe that the noise in the transcription adds much difficulty to models with lack of specialization in the Chinese, while LLMs that specialzes in Chinese could still detect the true text in the noisy transcription.
\subsection{Effect of scale}
\begin{figure}[htbp]
\centering
  \subfigure[Performances of Qwen1.5 series]{
    \begin{minipage}[t]{0.48\textwidth}
      \centering
      \includegraphics[width=\textwidth]{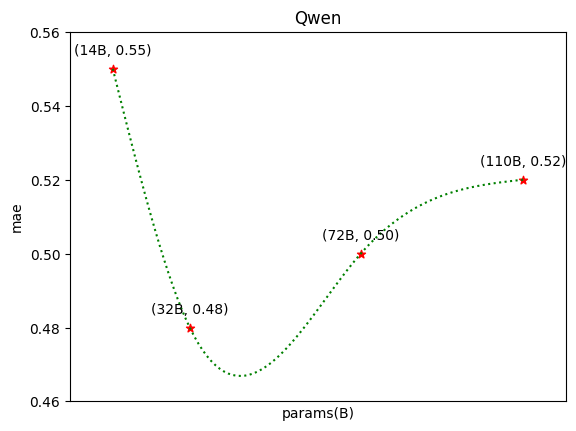}
      \label{fig:img1}
    \end{minipage}
  }
  \hfill
  \subfigure[Performances of Mistral series]{
    \begin{minipage}[t]{0.48\textwidth}
      \centering
      \includegraphics[width=\textwidth]{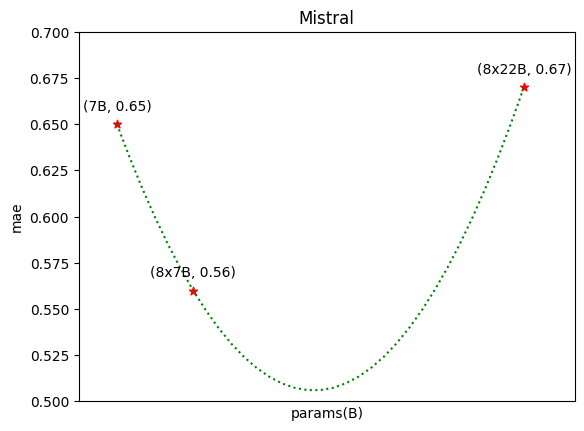}
      \label{fig:img2}
    \end{minipage}
  }
  \caption{Effect of scale. We compare the average MAE of 8 items of different sizes of two well-known open-source series models, Qwen1.5 and Mistral. }
  \label{fig:side_by_side}
\end{figure}

We find an interesting phenomenon: medium-sized models perform best in our task Figure \ref{fig:img1} \ref{fig:img2}. This forms an anti-scaling examples \cite{kaplan2020scaling}.
Why big-sized models fail? We attribute this to the fact that our task requires detecting abnormal through language. Big-sized models could more easily catch the noise in the transcription thus are misled to give higher scores. As for small-sized models, our task is too challenging for them. Many small-sized models even couldn't understand our task.

\section{More details for what effects LLMs' scoring of the eight ADOS items?}
\label{sec:what effect}
\begin{figure}
    \centering
    \includegraphics[width=1\linewidth]{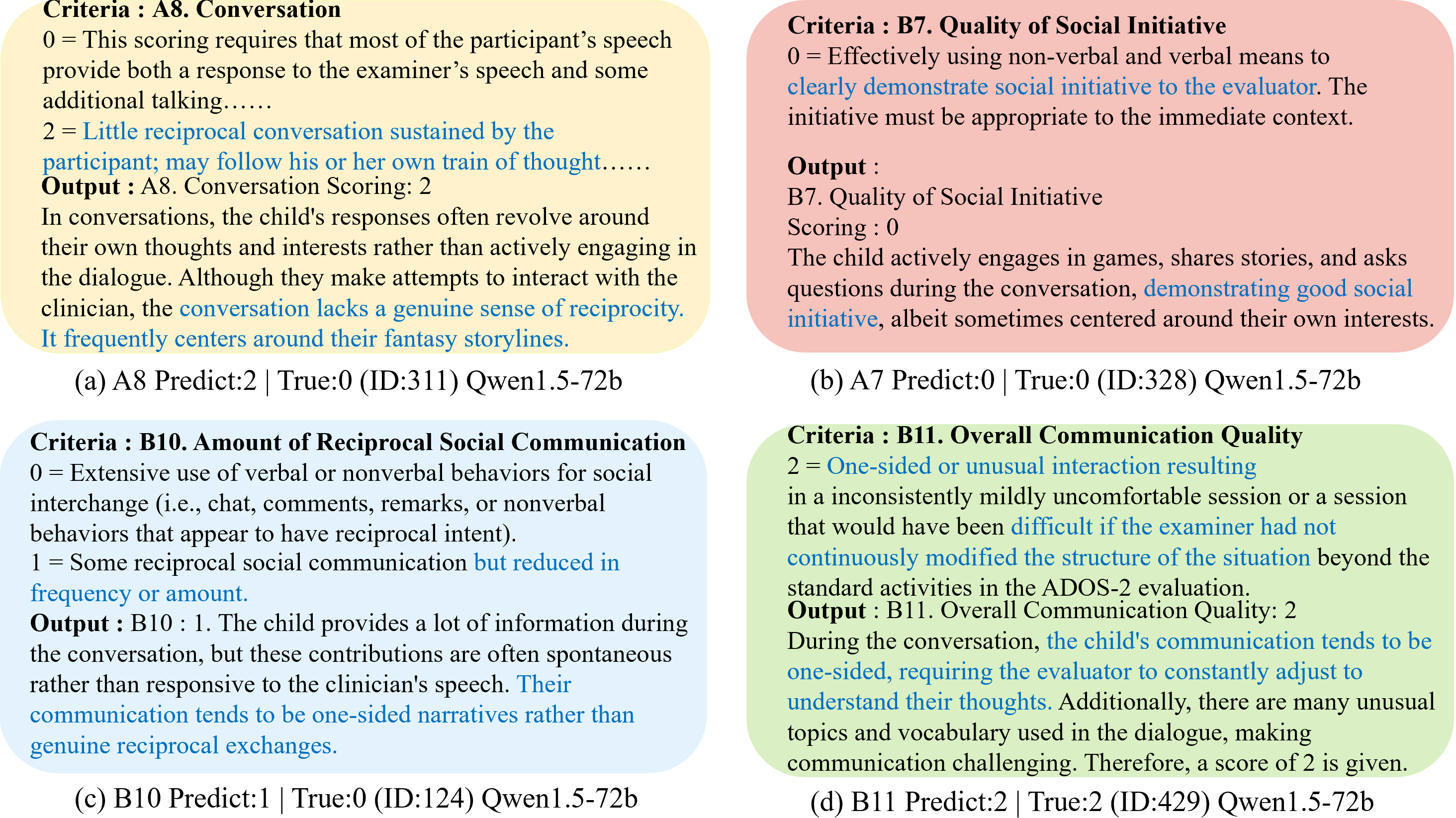}
    \caption{More case study for LLMs’preference Analysis. The \textcolor{blue}{blue} font in the figure indicates the basis for scoring.}
    \label{fig:appendix-case-study}
\end{figure}
\paragraph{A4}
corresponds to the stereotypical/special use of words or phrases, and the experimental results indicate that almost all LLMs perform poorly in this aspect. The reason behind this is that stereotypical/special use heavily relies on subjective judgment and often occurs sporadically in different dialogue contexts. In clinical practice, doctors can often determine stereotypical usage and assign a higher score based on a single interaction fragment and other non-verbal behavior. However, LLMs find it challenging to detect abnormalities from similar fragments just based on text, making it difficult for them to perceive the exception. As a result, LLMs can't align with the clinical doctor leading to this bias.
\paragraph{A7}
refers to event reporting, and LLMs perform not bad in this item. The purpose of this item is to identify communication between doctors and children regarding routine/unusual events. On one hand, LLMs need to detect relevant event fragments from the dialogue, and this type of task is widely present in the fine-tuning process of various LLMs. Therefore, LLMs are competent in this aspect. On the other hand, the ADOS scoring criteria for this item are described clearly, with distinct justifications for the three scores and corresponding examples provided. The combination of a clear prompt and LLMs' inherent capabilities enables the successful completion of this task.
\paragraph{A8}
corresponds to conversation and focuses on the fluency of social interactions and the level of word and phrase usage. LLMs struggle in this item because the evaluation of word and phrase usage is subjective and relies on the clinical experience of doctors. Moreover, assessing fluency and flexibility is more accurately done at the linguistic level, where aspects such as hesitations and interruptions are more evident in speech and tone than in written text. As shown in Figure \ref{fig:appendix-case-study}(a), LLMs perceive that from textual perspective, the child is immersed in their own world and lacks reciprocity during the conversation. However, the clinician's viewpoint is exactly the opposite. Although it is a conversation item, the nonverbal aspects of the clinical setting, such as eye contact or facial expressions, are certainly taken into consideration by the clinician, leading to significant discrepancies in scoring.
\paragraph{B4}
is a highly controversial item that involves the sharing of joy during interactions, and there is a significant variation in MAE among different LLMs in this item. On one hand, B4 involves the expression of happiness during interactions, which is not solely dependent on language. Factors such as body language, tone of voice, and facial expressions all influence the judgment of clinical doctors. LLMs, on the other hand, tend to exhibit more extreme scoring criteria based on the presented text. If there are no explicit scenes or semantic cues related to share enjoyment in the dialogue, the model tends to assign a higher score. However, once phrases like "hahaha, your story is very funny, i am happy." are detected, the model immediately assigns a lower score without considering the whole enjoyment of dialogues. Doctors, on the other hand, tend to have a more lenient and multidimensional approach to scoring this item, leading to the dramatic variation observed in the scores. On the other hand, The shared enjoyment in interactions can be regarded as a form of emotional analysis task, but it is more complex than traditional sentiment analysis due to its practical application in multi-turn dialogues where participants' emotional states undergo changes. Existing large models excel in simple sentiment analysis tasks, but they are outperformed by fine-tuned smaller models in complex tasks\cite{zhang2023sentiment}. Therefore, one possible reason for the underperformance of some large models on B4 is their insufficient capability to handle complex scenarios. We believe that part of the reason why some models achieve good results in this task is the utilization of prior statistical information in the prompt. This injection of prior statistics lowers the probabilities of tokens that were originally more likely in the model, causing the model to carefully consider extreme scoring and align its results with the doctor's thought process.
\paragraph{B7\&B9}
primarily assess social initiative and response and are summary items. LLMs can provide reasonably objective scores for the overall conversation situation. On one hand, this is attributed to the clear scoring descriptions for each score in the ADOS scoring criteria, with evident differences between the scores. On the other hand, in real assessment scenarios, doctors tend to carefully consider their scores for summary items, resulting in a more compromise scoring. LLMs, when evidence is not abundant, typically exhibit conservative scoring tendencies that align with doctors'preferences. As shown in Figure \ref{fig:appendix-case-study}(b), LLMs keenly observed the child's social initiative in games, storytelling, and questioning, leading to a score of 0.
\paragraph{B10\&B11}
assess the quality of reciprocal communication and evaluate the overall quality of communication. Unlike B7 and B9, the ADOS scoring criteria for these two items include descriptions of nonverbal behaviors. This makes LLMs less certain about their own scores when they only have text inputs. Additionally, the corresponding descriptions in the ADOS scoring criteria can be abstract, and unless a doctor is experienced and seasoned, they may struggle to fully understand the implications of their evaluations. This difficulty extends even more to LLMs. As shown in Figure \ref{fig:appendix-case-study}(c), LLMs perceive that the child's communication is spontaneous and one-sided rather than responsive to the clinician, resulting in fewer instances of reciprocal social communication, warranting a score of 1. However, the clinician's comprehensive observation from nonverbal and behavioral et al. aspects may subjectively lean towards a score of 0. Moreover, the boundary between scores 0 and 1 is not clearly defined, and relying solely on the phrase "reduced in frequency or amount" can lead to confusion for LLMs. In Figure \ref{fig:appendix-case-study} (d), B11 is a summary of the overall communication quality. LLMs perceive that the child's expressions during the conversation are one-sided, requiring the clinician to make huge efforts to sustain the dialogue. This aligns with the criteria for a score of 2, which is consistent with the clinician's judgment.

\begin{figure}
    \centering
    \includegraphics[width=1\linewidth]{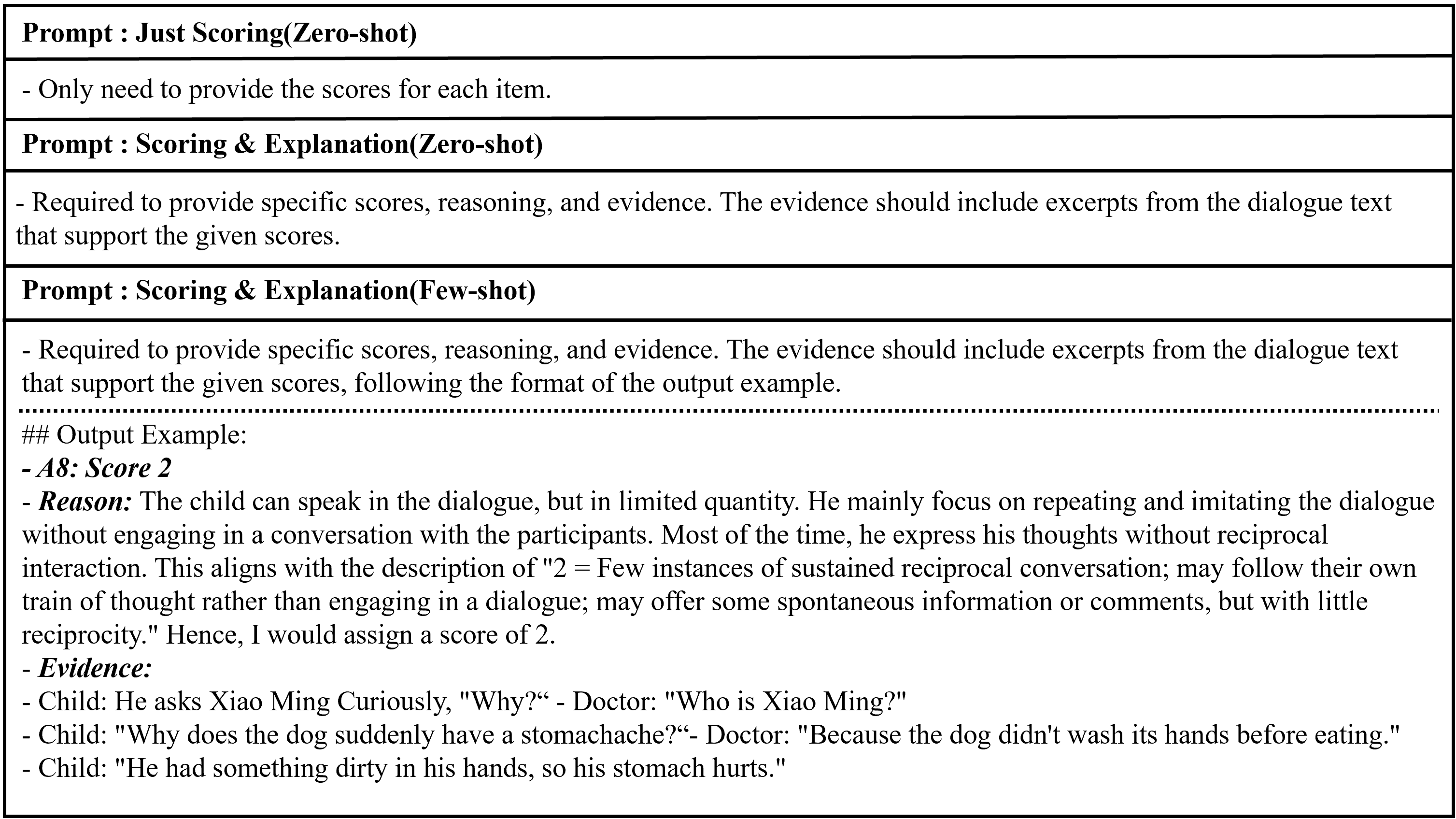}
    \caption{Optional Prompt For Zero/Few-shot Setting. This Figure illustrates the differences in three prompt settings. We only display the part that needs to be replaced for the \textcolor{blue}{-(optional)} shown in Figure \ref{fig:base prompt} while the left parts remain unchanged.}
    \label{fig:option prompt}
\end{figure}

\section{Zero/few-shot learning settings}
\label{sec:zs vs fs}
Moreover, we have devised three prompt settings, incorporating both zero-shot and few-shot settings, to investigate the hypotheses we have put forward in Method Section.
\begin{itemize}
    \item Only-Scoring (zero-shot):In the prompt, we require the model to provide scores for the eight items only, without the need for justifications.
    \item Scoring\&Explanation (zero-shot):In the prompt, we introduce CoT (Chain of thought)\cite{kojima2022large} to encourage the model to provide not only scores but also the reasoning and justifications behind those scores. This approach aims to showcase the model's thought process and achieve a more rigorous assessment of the scores.
    \item Scoring\&Explanation (few-shot):In the prompt, inspired by \cite{wei2022chain, zhang2023igniting}, we provide some few-shot examples that includes score, reasoning, and evidence, with the expectation that the model will generate similar outputs to gain more detail explanation.
\end{itemize}

Table \ref{tab:prompts-setting} refers to the comparison of results for several models in the \textbf{Only-Scoring} (\(os\)) and \textbf{Scoring\&Explanation (few-shot)} (\(fs\)) settings compared to the \textbf{Scoring\&Explanation (zero-shot)} setting. It is evident that in the only-scoring setting, all LLMs show a significant decrease in performance. The average MAE decreases by 2.84\%, the average F1 score for binary classification decreases by 10.61\%, and the average F1 score for ternary classification decreases by 13.66\%. This confirms our hypothesis that explanations aimed at better scoring. In the few-shot setting, there is an issue with Kimi's refusal to score certain items, which may be attributed to the sensitive settings. On the other hand, GLM4 shows a slight performance improvement while Qwen1.5 and Yi-34b show the performance decline. This once again confirms that the few-shot setting does not necessarily lead to significant performance improvements in novel tasks which is never used in pre-training and fine-tuning for LLMs\cite{li2024task}. Therefore, in the main text, we opted to use the \textbf{Scoring\&Explanation (zero-shot)} prompt setting as the In-context Enhancement Prompt which is shown in Figure \ref{fig:base prompt}.

\begin{table}[]
\caption{Performances Comparison of  Different Prompts Setting. This table displays the performance comparison results between the Scoring\&Explanation(zero-shot) setting, where \textit{os} refers to the Only-Scoring(zero-shot) setting and \textit{fs} refers to the Scoring\&Explanation(few-shot) setting. The \textcolor{blue}{blue} indicate a decrease in performance compared to the Scoring\&Explanation(zero-shot) setting, while the \textcolor{red}{red} indicate an improvement in performance.}
\label{tab:prompts-setting}
% \resizebox{\textwidth}{!}{
%
\centering
\begin{tabular}{@{}clllllll@{}}
\toprule
\multicolumn{1}{l}{\diagbox{metric}{model}} &
  \multicolumn{1}{c}{Kimi} &
  \multicolumn{1}{c}{qwen1.5-32b} &
  \multicolumn{1}{c}{Yi-34b} &
  \multicolumn{1}{c}{glm4} &
    \multicolumn{1}{c}{Avg} 
  \\\midrule
\textbf{\(Avg-MAE_{os}\)} &
  \textcolor{blue}{\(\downarrow\) -0.0259} &
  \textcolor{blue}{\(\downarrow\) -0.0292} &
  \textcolor{blue}{\(\downarrow\) -0.0259} &
  \textcolor{blue}{\(\downarrow\) -0.0324} &
  \textcolor{blue}{\(\downarrow\) -0.0284} \\
\textbf{\(F1-2_{os}\)} &
  \textcolor{blue}{\(\downarrow\) -0.0880} &
  \textcolor{blue}{\(\downarrow\) -0.1681} &
  \textcolor{blue}{\(\downarrow\) -0.0127} &
  \textcolor{blue}{\(\downarrow\) -0.1556} &
  \textcolor{blue}{\(\downarrow\) -0.1061} \\

\textbf{\(F1-3_{os}\)} &
    \textcolor{blue}{\(\downarrow\) -0.1641} &
    \textcolor{blue}{\(\downarrow\) -0.1015} &
    \textcolor{blue}{\(\downarrow\) -0.0727} &
    \textcolor{blue}{\(\downarrow\) -0.2080} &
    \textcolor{blue}{\(\downarrow\) -0.1366} \\
\midrule
\textbf{\(Avg-MAE_{fs}\)} &
  \textbackslash &
  \textcolor{blue}{\(\downarrow\) -0.0178} &
  \textcolor{blue}{\(\downarrow\) -0.1039} &
  \textcolor{red}{\(\uparrow\) +0.0293} &
  \textcolor{blue}{\(\downarrow\) -0.0326} \\
\textbf{\(F1-2_{fs}\)} &
   \textbackslash &
   \textcolor{blue}{\(\downarrow\) -0.1227} &
   \textcolor{blue}{\(\downarrow\) -0.0853} &
   \textcolor{red}{\(\uparrow\) +0.0914} &
  \textcolor{blue}{\(\downarrow\) -0.0389} \\ 
\textbf{\(F1-3_{fs}\)} &
   \textbackslash &
   \textcolor{blue}{\(\downarrow\) -0.0931} &
   \textcolor{blue}{\(\downarrow\) -0.0673} &
   \textcolor{red}{\(\uparrow\) +0.0704} &
  \textcolor{blue}{\(\downarrow\) -0.0300} \\ 
  \bottomrule
\end{tabular}
\end{table}

\end{document}